\documentclass[12pt,letterpaper]{article}
\usepackage{jheppub}
\usepackage{aas_macros}
\usepackage{bbm}
\usepackage{graphicx} 
\usepackage{caption}
\usepackage{subcaption}
\usepackage[toc,page]{appendix}

\newcommand{\be}{\begin{equation}}
\newcommand{\ee}{\end{equation}}
\newcommand{\ev}[1]{\left\langle#1\right\rangle}

\title{The Bayesian Second Law of Thermodynamics}
\preprint{CALT-TH-2015-016}
\author[a]{Anthony Bartolotta,}
\affiliation[a]{Walter Burke Institute for Theoretical Physics,\\
California Institute of Technology, Pasadena, CA 91125, U.S.A.}
\emailAdd{abartolo@caltech.edu}

\author[a]{Sean M. Carroll,}
\emailAdd{seancarroll@gmail.com}

\author[b]{Stefan Leichenauer,}
\affiliation[b]{Center for Theoretical Physics and Department of Physics,\\
 University of California, Berkeley, CA 94720, U.S.A.}
\emailAdd{sleichen@berkeley.edu}

\author[a]{and Jason Pollack}
\emailAdd{jpollack@caltech.edu}

\abstract{We derive a generalization of the Second Law of Thermodynamics that uses Bayesian updates to explicitly incorporate the effects of a measurement of a system at some point in its evolution. By allowing an experimenter's knowledge to be updated by the measurement process, this formulation resolves a tension between the fact that the entropy of a statistical system can sometimes fluctuate downward and the information-theoretic idea that knowledge of a stochastically-evolving system degrades over time. The Bayesian Second Law can be written as $\Delta H(\rho_m, \rho) + \langle \mathcal{Q}\rangle_{F|m}\geq 0$, where $\Delta H(\rho_m, \rho)$ is the change in the cross entropy between the original phase-space probability distribution $\rho$ and the measurement-updated distribution $\rho_m$, and $\langle \mathcal{Q}\rangle_{F|m}$ is the expectation value of a generalized heat flow out of the system. We also derive refined versions of the Second Law that bound the entropy increase from below by a non-negative number, as well as Bayesian versions of the Jarzynski equality. We demonstrate the formalism using simple analytical and numerical examples.
}

\begin{document}
\maketitle

\vfill\eject

\section{Introduction}

The Second Law of Thermodynamics encapsulates one of the most important facts about the macroscopic world: entropy increases over time. There are, however, a number of different ways to define ``entropy,'' and corresponding controversies over how to best understand the Second Law. In this paper we offer a formulation of the Second Law that helps to resolve some of the tension between different approaches, by explicitly including the effects of the measurement process on our knowledge of the state of the system. This Bayesian Second Law (BSL) provides a new tool for analyzing the evolution of statistical systems, especially for small numbers of particles and short times, where downward fluctuations in entropy can be important.

One way to think about entropy and the Second Law, due to Boltzmann, coarse-grains the phase space $\Gamma$ of a system into macrostates. The entropy of a microstate $x$ is then given by $S = \log\Omega_x$, where $\Omega_x$ is the volume of the macrostate to which $x$ belongs. (Throughout this paper we set Boltzmann's constant $k_B$ equal to unity.) The coarse-graining itself is subjective, but once it is fixed there is a definite entropy objectively associated with each microstate. Assuming that the system starts in a low-entropy state (the ``Past Hypothesis''), the Second Law simply reflects the fact that undirected evolution is likely to take the state into ever-larger macrostates: there are more ways to be high-entropy than to be low-entropy. The Second Law is statistical, in the sense that random fluctuations into lower-entropy states, while rare, are certainly possible. In many contexts of interest to modern science, from nanoscale physics to biology, these fluctuations are of crucial importance, and the study of ``fluctuation theorems'' has garnered considerable attention in recent years \cite{Jarzynski:1997, crooks1998nonequilibrium, Crooks:1999, evans2002fluctuation, jarzynski2011equalities, sagawa2012nonequilibrium, england2013statistical}.

Another perspective on entropy, associated with Gibbs in statistical mechanics and Shannon \cite{Shannon:1948zz} in the context of information theory, starts with a normalized probability distribution $\rho(x)$ on phase space, and defines the entropy as $S=-\int dx\,\rho(x)\log\rho(x)$. In contrast with the Boltzmann formulation, in this version the entropy characterizes the state of our knowledge of the system, rather than representing an objective fact about the system itself. The more spread-out and uncertain a distribution is, the higher its entropy. The Second Law, in this view, represents the influence of stochastic dynamics on the evolution of the system, for example due to interactions with a heat bath, under the influence of which we know less and less about the microstate of the system as time passes. 

For many purposes, the Gibbs/Shannon formulation of entropy and the Second Law is more convenient to use than the Boltzmann formulation. However, it raises a puzzle: how can entropy ever fluctuate downward? In an isolated system evolving according to Hamiltonian dynamics, the Gibbs entropy is strictly constant, rather than increasing; for a system coupled to a heat bath with no net energy transfer, it tends to monotonically increase, asymptoting to a maximum equilibrium value. Ultimately this is because the Gibbs entropy characterizes our knowledge of the microstate of the system, which only diminishes with time.\footnote{Boltzmann himself also studied a similar formulation of entropy, which he used to prove his $H$-theorem. The difference is that the $H$-functional represents $N$ particles in one 6-dimensional single-particle phase space, rather than in a $6N$-dimensional multi-particle phase space. This is not a full representation of the system, as it throws away information about correlations between particles. The corresponding dynamics are not reversible, and entropy increases \cite{Zeh}.}

We can, of course, actually \emph{observe} the system; if we do so, we will (extremely) occasionally notice that it has fluctuated into what we would characterize as a low-entropy state from Boltzmann's perspective. The air in a room could fluctuate into one corner, for example, or a cool glass of water could evolve into a warm glass of water containing an ice cube. To reconcile this real physical possibility with an information-centric understanding of entropy, we need to explicitly account for the impact of the act of measurement on our knowledge of the system. This is the task of Bayesian analysis, which shows us how to update probability distributions in the face of new information \cite{Bayes:1764vd,Laplace}. Since the advent of Maxwell's demon, measurement in the context of statistical mechanics has been explored extensively \cite{RevModPhys.81.1}. This has resulted in a body of literature linking information-theoretic quantities to thermodynamic variables \cite{parrondo2015thermodynamics, PhysRevA.39.5378}. However, such analyses only examine the impact of measurement at the point in time when it is performed. In the present work, we observe that such measurements also contain information about the state of the system at earlier points in time that are hitherto unaccounted for. This results in novel modifications of the Second Law.

The setup we consider consists of a classical system coupled to an environment. The dynamics of the system are stochastic, governed by transition probabilities, either due to intrinsic randomness in the behavior of the system or to the unpredictable influence of the environment. An experimental protocol is determined by a set of time-dependent parameters, which may be thought of as macroscopic features (such as the location of a piston) controlled by the experimenter. The experimenter's initial knowledge of the system is characterized by some probability distribution; as the system is evolved under the protocol for some period of time, this probability distribution also evolves. At the end of the experiment, the experimenter performs a measurement. Bayes's Theorem tells us how to update our estimates about the system based on the outcome of the measurement; in particular, we can use the measurement outcome to update the final probability distribution, but also to update the \emph{initial} distribution. The BSL is a relation between the original (non-updated) distributions, the updated distributions, and a generalized heat transfer between the system and the environment. 

The Second Law contains information about irreversibility; a crucial role in our analysis is played by the relationship between transition probabilities forward in time and ``reversed'' probabilities backward in time. Consider a situation in which the system in question is an egg, and the experiment consists of holding the egg up and dropping it. To be precise, the experimental protocol, which we will call the ``forward" protocol, is for the experimenter to hold the egg in the palm of her open hand, and then to turn her hand over after a specified amount of time. The initial probability distribution for the particles that make up the egg is one that corresponds to an intact egg in the experimenter's hand. With overwhelming probability the forward protocol applied to this initial state will result in an egg on the floor, broken.

This experiment is clearly of the irreversible type, but we should be careful about why and how it is irreversible. If we reversed the velocities of every particle in the universe, then time would run backward and the egg would reconstitute itself and fly back up into the experimenter's hand. This sort of fundamental reversibility is not what concerns us. For us, irreversibility means that there are dissipative losses to the environment: in particular, there are losses of \emph{information} as the state of the system interacts with that of the environment. This information loss is what characterizes irreversibility. From the theoretical viewpoint, we should ask what would happen if all of the velocities of the broken egg particles were instantaneously reversed, leaving the environment alone. Again with overwhelming probability, the egg would remain broken on the floor. To make sure the time-dependent actions of the experimenter do not affect this conclusion, we should also instruct the experimenter to run her experiment in reverse: she should begin with her palm facing downward while the egg is broken on the floor, and then turn it upward after a certain amount of time. In this example, the effect of reversing the experimental procedure is negligible; the probability that the egg will reassemble itself and hop up into her hand is not zero, but it is extremely small.

The generalization beyond the egg dropping experiment is clear. We have a system and an environment, and an experimenter who executes a forward protocol, which means a macroscopic time-dependent influence on the dynamics of the system. The environmental interactions with the system are deterministic but unknown to the experimenter, and so the system evolves stochastically from her point of view. She assigns probabilities to trajectories the system might take through phase space. We will call these the ``forward" probabilities. To isolate potential irreversibility in the system, we consider reversing all of the velocities of the system's particles in its final state, and then executing the ``reverse" protocol, which is just the forward protocol backward. The environment still interacts in an unknown way, so the system again evolves stochastically. The probabilities that the experimenter assigns to trajectories in this reversed setup are called the reverse probabilities.

To get precise versions of the Second Law, we will consider a particular information-theoretic measure of the difference between the forward and reverse probabilities, known as the relative entropy or Kullback-Leibler divergence \cite{kullback1951}. The relative entropy of two probability distributions is always non-negative, and vanishes if and only if the two distributions are identical. The relative entropy of the forward and reverse probability distributions on phase space trajectories is a measure of the irreversibility of the system, and the non-negativity of that relative entropy is a precise version of the Second Law. 

The inclusion of Bayesian updates as the result of an observation at the end of the protocol leads to the Bayesian Second Law. The BSL can be written in several ways, one of which is:
\be
  \Delta H(\rho_m, \rho) + \langle \mathcal{Q}\rangle_{F|m}\geq 0.
\label{bsl1}
\ee
Here, $\rho$ is the probability distribution without updating, and $\rho_m$ is the updated distribution after obtaining measurement outcome $m$. $H = -\int \rho_m \log \rho$ is the cross entropy between the two distributions. The cross entropy is the sum of the entropy of $\rho_m$ and the relative entropy of $\rho_m$ with respect to $\rho$; it can be thought of as the average amount we would learn about the system by being told its precise microstate, if we thought it was in one distribution (the original $\rho$) but it was actually in another (the updated $\rho_m$). Like the ordinary entropy, this is a measure of uncertainty: the more information contained in the (unknown) microstate, the greater the uncertainty. However, the cross entropy corrects for our false impression of the distribution. The difference in the cross entropy between the initial and final times is $\Delta H$, and $\langle \mathcal{Q}\rangle_{F|m}$ is the expectation value of a generalized heat transfer between the system and the environment, which contains information about the irreversibility of the system's dynamics. Thus, at zero heat transfer, the BSL expresses the fact that our uncertainty about the system is larger at the time of measurement, even after accounting for the measurement outcome.

The relative entropy is not only non-negative, it is monotonic: if we apply a stochastic (probability-conserving) operator to any two distributions, the relative entropy between them stays constant or decreases. We can use this fact to prove refined versions of both the ordinary and Bayesian Second Laws, obtaining a tighter bound than zero to the expected entropy change plus heat transfer. This new lower bound is the relative entropy between the initial probability distribution and one that has been cycled through forward and reverse evolution, and therefore characterizes the amount of irreversibility in the evolution. 

We also apply our implementation of Bayesian updating to the Jarzynski equality, which relates the expected value of the work performed during a process to the change in free energy between the initial and final states. Lastly, we illustrate the BSL in the context of some simple models. These include deriving Boltzmann's version of the Second Law within our formalism, and studying the numerical evolution of a randomly driven harmonic oscillator.

\section{Setup}

\subsection{The System and Evolution Probabilities}

We are primarily concerned with dynamical systems that undergo non-deterministic evolution, typically due to interactions with an environment about which the experimenter has no detailed knowledge. The effect of the unknown environment is to induce effectively stochastic evolution on the system; as such, we can only describe the state and subsequent time evolution of the system probabilistically \cite{seifert2008stochastic}. We are considering classical mechanics, where probabilities only arise due to the ignorance of the experimenter, including ignorance of the state of the environment. Analogous equations would apply more generally to truly stochastic systems, or to stochastic models of dynamical systems. 

The state of the system at time $t$ is therefore a random variable $X_t$ taking values in a space of states $\Gamma$. We will refer to $\Gamma$ as ``phase space,'' as if it were a conventional Hamiltonian system, although the equations apply equally well to model systems with discrete state spaces.
Because the evolution is non-deterministic, we can only give a probability that the system is in state $x$ at time $t$, which we write as $P(X_t = x)$. This is a true probability in the discrete case; in the continuous case it is more properly a probability density  that should be integrated over a finite region of $\Gamma$ to obtain a probability, but we generally will not draw this distinction explicitly. For notational convenience, we will often write this probability as a distribution function,
\be
\label{eq:def-rho_t}
\rho_t(x) \equiv P(X_t = x),
\ee
which is normalized so that $\int \rho_t(x)\,dx=1$.

The experimenter has two roles: to manipulate a set of external control parameters defining the experimental protocol, and to perform  measurements on the system. All measurements are assumed to be ``ideal''; that is, the act of measuring any given property of the system is assumed to induce no backreaction on its state, and we do not track the statistical properties of the measuring device. 

We will primarily be studying experiments that take place over a fixed time interval $\tau$. The experimental protocol is fully specified by the history of a set of external control parameters that can change over this time interval, $\lambda_i ( t )$. The control parameters $\lambda_i$ specify the behavior of various external potentials acting on the system, such as the volume of a container or the frequency of optical tweezers. We will refer to the set $\lambda(t) = \{\lambda_i(t)\}$ of control parameters as functions of time as the ``forward protocol.'' 

The forward protocol and the dynamics of the system together determine the forward transition function, $\pi_F$, which tells us the probability that the system evolves from an initial state $x$ at $t=0$ to a final state $x'$ at $t=\tau$:
\be
  \pi_{F}(x \to x' ) \equiv P(X_\tau = x' | X_0 = x ; \lambda \left( t\right) ).
 \label{eq:pi_F}
\ee
The transition function $\pi_F$ is a conditional probability, normalized so that the system ends up somewhere with probability one:
\be
  \int \pi_{F}(x \to x') dx' = 1.
  \label{eq:pi-normalization}
\ee
The forward transition function evolves the initial distribution to the final distribution,
\be
  \rho_\tau(x') = \int dx\, \rho_0(x) \pi_F(x\rightarrow x').
  \label{eq:rho-evolve}
\ee

A central role will be played by the joint probability that the system begins at $x$ and ends up a time $\tau$ later at $x'$,\footnote{Here and below we will mostly omit the dependence on the control parameters $\lambda(t)$ from the notation for brevity. They will return in Section~\ref{sec-reversal} when we discuss time-reversed experiments.}
\be
P_F(x,x') \equiv P(X_0 = x, X_\tau = x') = \rho_0(x) \pi_F(x \to x'),
\label{eq-jointdef}
\ee
which is normalized so that $\int P(x,x')\,dx dx' = 1$.
By summing the joint probability over $x$ or $x'$ we obtain the distribution functions $\rho_\tau(x')$ or $\rho_0(x)$, respectively:
\begin{align}
\rho_\tau(x') &= \int P_F(x,x') dx,\cr
\rho_0(x) &= \int P_F(x,x') dx'.
\label{eq:rho-from-P}
\end{align}

We close this subsection with a brief digression on the probabilities of phase-space trajectories. The rules of conditional probability allow us to break up the transition functions based on subdivisions of the time interval $[0,\tau]$. For the special case of a Markov process, we have the identity
\begin{align}
\pi_F(x\to x') =& \int [dx] \,P(X_\tau = x'| X_{t_N} = x_N) \cr
& \times P(X_{t_N} = x_N | X_{t_{N-1}} = x_{N-1})\cdots P(X_{t_1} = x_1 | X_0 = x),
\end{align}
where $[dx]$ is the product of all the $dx_k$ and we choose $t_k = k\tau/(N+1)$. This is familiar as a discretization of the path integral, and in the continuum limit we would write
\be
\pi_F(x\to x') = \int_{x(0) = x}^{x(\tau) = x'} \mathcal{D}x(t) \,\pi_F[x(t)].
\ee
The functional $\pi_F[x(t)]$ is a probability density on the space of trajectories with fixed initial position, but with the final position free. To get a probability density on the space of trajectories with two free endpoints, we just have to multiply $\pi_F[x(t)]$ by the initial distribution $\rho_0(x)$. The result, which we call $P_F[x(t)]$, is the path-space version of the joint distribution $P_F(x,x')$. We will not make heavy use of these path-space quantities below, but the formal manipulations we make with the ordinary transition function and joint distribution can be repeated exactly with the path-space distributions, and occasionally we will comment on the path-space versions of our results.

\subsection{Measurement and Bayesian Updating}

The probability density on phase space can also change through Bayesian updates when a measurement is made: the experimenter modifies her probabilities to account for the new information. We will restrict ourselves to measurements performed at time $\tau$, the end of the experiment, though it is simple to extend the results to more general measurement protocols. The measurement outcome is a random variable $M$ that only depends on the state of the system at time $\tau$, not on the prior history of the system. The measurement is then characterized by the function
\begin{align}
\label{eq:Measurement}
P(m | x') & \equiv  P(M=m | X_\tau = x') \\
& =  \text{probability of measurement outcome $m$ given state $x'$ at time $\tau$}.\nonumber
\end{align}
The updated phase space distribution at time $\tau$ is obtained by Bayes's rule, which in this case takes the form
\be
\rho_{\tau|m}(x') \equiv P(X_\tau = x' | M = m)  = \frac{P(m|x')}{P(m)}\rho_\tau(x').
\label{updated-final}
\ee
Here the denominator is $P(m)\equiv\int P(m|y')\rho_\tau(y')dy'$, and serves as a normalization factor. 

\begin{figure}[t]
	\begin{center}
	\includegraphics[width=.75\textwidth]{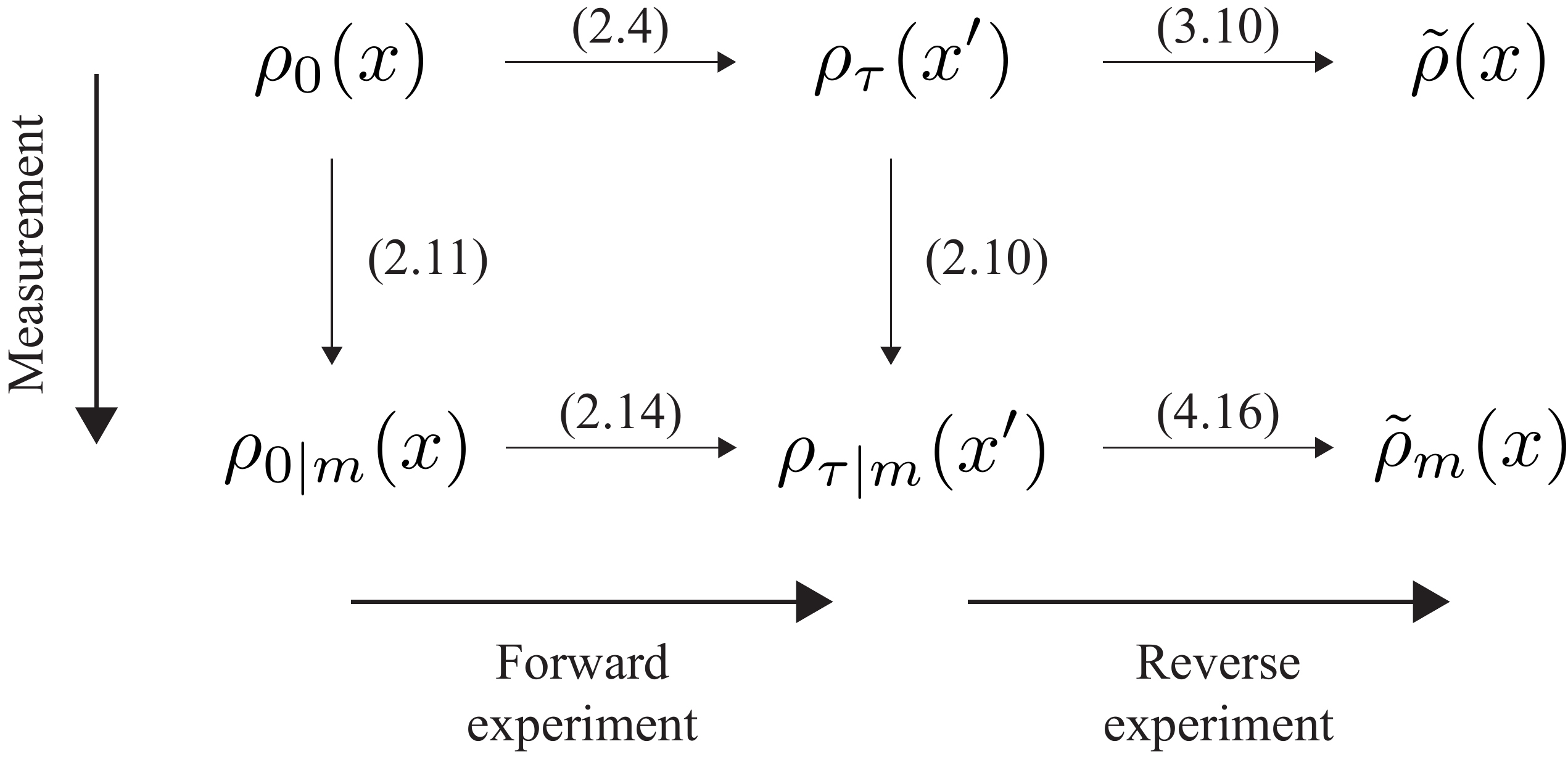}
	\end{center}
	\caption{Relationships between the various distribution functions we define: the original distribution $\rho_0(x)$, its time-evolved version $\rho_\tau(x')$, their corresponding Bayesian-updated versions $\rho_{0|m}(x)$ and $\rho_{\tau|m}(x')$, and the cycled distributions $\tilde\rho(x)$ and $\tilde\rho_{m}(x)$ discussed in Sections~\ref{refinedsecondlaw} and \ref{sec-refinedBSL}. Equation numbers refer to where the distributions are related to each other.}
	\label{fig:rho}
\end{figure}

If we know the transition function, we can also update the phase space distribution at any other time based on the measurement outcome at time $\tau$. Below we will make use of the updated initial distribution:
\be
\rho_{0|m}(x) \equiv P(X_0 = x | M=m) = \frac{\rho_0(x) \int dx'~\pi_F(x\to x')P(m|x')}{P(m)}.
\label{updated-initial}
\ee
This reflects our best information about the initial state of the system given the outcome of the experiment; $\rho_{0|m}(x)$ is the probability, given the original distribution $\rho_0(x)$ and the measurement outcome $m$ at time $t=\tau$, that the system was in state $x$ at time $t=0$. For example, we may initially be ignorant about the value of an exactly conserved quantity. If we measure it at the end of the experiment then we know that it had to have the same value at the start; this could mean a big difference between $\rho_0$ and $\rho_{0|m}$, though often the effects will be more subtle. The various distribution functions we work with are summarized in Figure~\ref{fig:rho} and listed in Table \ref{tab:distributions}.

\begin{table}[t]
\begin{center} \begin{tabular}{  l  l  c  } \hline 
Distribution & Name & Definition \\ \hline 
$\rho_{0}(x)$ & Initial Distribution & \ref{eq:def-rho_t} \\
$\pi_{F}(x\rightarrow x')$ & Forward Transition Function & \ref{eq:pi_F} \\
$\rho_{\tau}(x')$ & Final Distribution & \ref{eq:rho-evolve} \\
$P_{F}(x,x')$ & Joint Forward Distribution & \ref{eq-jointdef} \\ \hline 
$P(m|x)$ & Measurement Function & \ref{eq:Measurement} \\
$\rho_{\tau|m}(x')$ & Updated Final Distribution & \ref{updated-final} \\
$\rho_{0|m}(x)$ & Updated Initial Distribution & \ref{updated-initial} \\
$\pi_{F|m}(x\rightarrow x')$ & Updated Forward Transition Function & \ref{eq:pi_Fm} \\
$P_{F|m}(x,x')$ & Updated Joint Forward Distribution & \ref{eq:P_F|m} \\ \hline
$\pi_{R}(\overline{x'}\rightarrow \overline{x})$ & Reverse Transition Function & \ref{eq:pi_R} \\
$P_{R}(x,x')$ & Joint Reverse Distribution & \ref{eq:P_R} \\
$P_{R|m}(x,x')$ & Updated Joint Reverse Distribution & \ref{eq:P_R} \\
$\tilde{\rho}(x)$ & Cycled Distribution & \ref{eq:cycled-distribution} \\
$\tilde{\rho}_{m}(x)$ & Updated Cycled Distribution & \ref{eq:updated-cycled-distribution} \\ \hline
\end{tabular} \end{center}
\caption{List of named probability distributions and their defining equations. These are grouped according to whether they are updated and/or time-reversed.}
\label{tab:distributions}
\end{table}

Finally, we can update the forward transition functions,
\be
\pi_{F|m}(x\to x') \equiv P(X_\tau = x' | X_0 = x, M=m) = \frac{\pi_F(x \to x')P(m|x')}{\int dy' ~\pi_F(x \to y')P(m|y')},
\label{eq:pi_Fm}
\ee
and the joint distributions,
\be
P_{F|m}(x,x')  \equiv P(X_0 = x, X_\tau = x' | M=m) = \frac{P(m|x')}{P(m)}P_F(x,x') =  \rho_{0|m}(x) \pi_{F|m}(x \to x'), \label{eq:P_F|m}
\ee
based on the measurement outcome. As we would expect, the updated transition function evolves the updated distribution from the initial to the final time:
\be
  \rho_{\tau|m}(x') = \int dx\, \rho_{0|m}(x)\pi_{F|m}(x\rightarrow x').
\label{updated-final-pi}
\ee
It may seem odd to update the transition functions based on measurements, since in principle the original transition functions were completely determined by the stochastic dynamics of the system and this is a desirable property that one would like to preserve. For this reason, the unupdated transition functions will play a special role below, while the updated ones are only used as an intermediate quantity in algebraic manipulations.

\begin{figure}[h]
	\begin{center}
	\includegraphics[width=.95\textwidth]{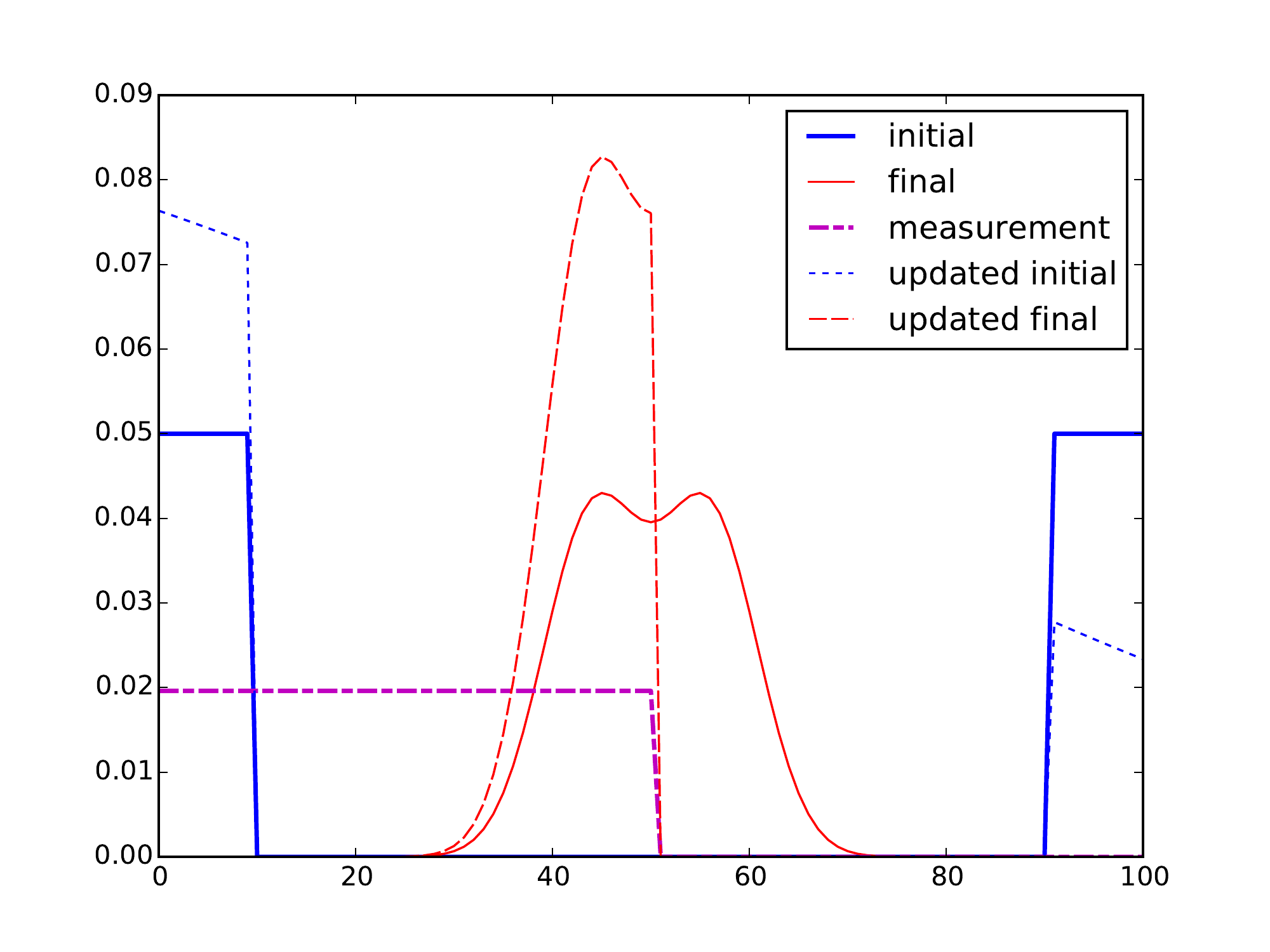}
	\end{center}
	\caption{The various distribution functions illustrated within a toy model of 100 independent spins with a fixed chance of flipping at every timestep. The distributions are normalized functions on the space of the total number $x$ of up-spins. We consider an initial distribution (thick solid blue line) that is equally split between the intervals $x<10$ and $90 < x$. The system is evolved for enough time to come close to equilibrium but not quite reach it, as shown by the final distribution (thin solid red line). A measurement is performed, revealing that less than half of the spins are up (dot-dashed purple line). We can therefore update the post-measurement final distribution (dashed red line). The corresponding updated initial distribution (dotted blue line) is similar to the original initial distribution, but with a boost at low $x$ and a decrease at high $x$.}
	\label{fig:spins}
\end{figure}

To illustrate these definitions, consider a simple toy model: a collection of $N$ independent classical spins, each of which has a fixed probability to flip its state at each timestep. In this model it is most intuitive to work with a distribution function defined on macrostates (total number of up spins) rather than on microstates (ordered sequences of up/down spins). 

The distribution functions relevant to our analysis are illustrated for this toy model with $N=100$ spins in Fig.~\ref{fig:spins}. To make the effects of evolution and updating most clear, we start with a bimodal initial distribution $\rho_0(x)$, uniform on the intervals $0 \leq x < 10$ and $90< x \leq 100$. The system is evolved for a short time $\tau$, not long enough to attain the equilibrium distribution, which would be a binomial centered at $x = N/2 = 50$. The final distribution $\rho_\tau(x')$  therefore has two small peaks just above and below $x'=50$. We then perform a measurement, which simply asks whether most of the spins are up or down, obtaining the answer ``mostly down.'' This corresponds to a measurement function 
\be
P(m | x) = \begin{cases}
1~{\rm if}~x\leq 50,\\
0~{\rm if}~x > 50.
\end{cases}
\ee
In Fig.~\ref{fig:spins} we have plotted the normalized version $P(m|x)/P(m)$. From this we can construct the updated final and updated initial distributions, using (\ref{updated-final}) and (\ref{updated-initial}). The updated final distribution is just the left half of the non-updated final distribution, suitably renormalized. The updated initial distribution is a re-weighted version of the non-updated initial distribution, indicating that there is a greater probability for the system to have started with very few up spins (which makes sense, since our final measurement found that the spins were mostly down). This toy model does not have especially intricate dynamics, but it suffices to show how our evolution-and-updating procedure works.

\subsection{The Reverse Protocol and Time Reversal}\label{sec-reversal}

The Second Law contains information about the irreversibility of the time-evolution of the system, so to derive it we need to specify procedures to time-reverse both states and dynamics. Specifically, we will define an effectively ``time-reversed" experiment that we can perform whose results can be compared to the time-forward experiment. As discussed in the Introduction, the point here is not to literally reverse the flow of time upon completion of the time-forward experiment (which would just undo the experiment), but to isolate the effects of dissipative processes, like friction, which result from complicated interactions with the environment.

For a state $x$, we denote by $\overline{x}$ the time-reversed state. In a ballistic model of particles, $\overline{x}$ is just the same as $x$ with all of the particle velocities reversed. We are only talking about the velocities of the particles that make up the system, not the environment. In practice, an experimenter is not able to control the individual velocities of all of the particles in the system, so it may seem pointless to talk about reversing them. It will often be possible, however, to set up a time-reversed probability distribution $\overline{\rho}(x) \equiv \rho(\overline{x})$ given some procedure for setting up $\rho(x)$. For instance, if the system has a Maxwellian distribution of velocities with zero center-of-mass motion, then the probability distribution on phase space is actually time-reversal invariant.

Time reversal of dynamics is simpler, primarily because we have only limited experimental control over them. The system will have its own internal dynamics, it will interact with the environment, and it will be influenced by the experimenter. In a real experiment, it is only the influence of the experimenter that we are able to control, so our notion of time reversal for the dynamics is phrased purely in terms of the way the experimenter decides to influence the system. The experimenter influences the system in a (potentially) time-dependent way by following an experimental protocol, $\lambda (t)$, which we have called the ``forward protocol." The forward protocol is a sequence of instructions to carry out using some given apparatus while the experiment is happening. We therefore define a ``reverse protocol," which simply calls for the experimenter to execute the instructions backward. In practice, that involves time-reversing the control parameters ({\it e.g.}, reversing macroscopic momenta and magnetic fields) and running them backwards in time, sending
\be
\lambda_i(t) \rightarrow \bar\lambda_i(\tau-t).
\ee
For simplicity we will generally assume that the control parameters are individually invariant under time-reversal, so we won't distinguish between $\lambda$ and $\bar\lambda$. The non-trivial aspect of the reverse protocol is then simply exchanging $t$ with $\tau-t$. If the control parameters are time-independent for the forward protocol, then there will be no difference between the forward and reverse protocols. This kind of experiment involves setting up the initial state of the system and then just waiting for a time $\tau$ before making measurements.

Recall that the transition functions $\pi_F$ for the system were defined assuming the experimenter was following the forward protocol. The reverse protocol is associated with a set of reverse transition functions $\pi_R$. We define $\pi_R$ in analogy with \eqref{eq:pi_F} as 
\be
\label{eq:pi_R}
\pi_{R}(\overline {x'} \to \overline x ) \equiv P(X_\tau = \overline x | X_0 = \overline{x'} ; \lambda \left( \tau - t\right) ),
\ee
normalized as usual so that $\int \pi_{R}(\overline{x'} \to \overline{x}) \, d\overline{x} = 1$.

We will also need a time-reversed version of the joint distribution $P_F$. As before, let $\rho_0(x)$ denote the initial distribution, and let $\rho_{\tau|m}(x)$ and $\rho_{\tau}(x)$ denote the distributions at time $\tau$ after following the forward protocol with and without Bayesian updates due to measurement, respectively. Then, following \eqref{eq-jointdef} and \eqref{eq:P_F|m}, define
\begin{align}
\label{eq:P_R}
P_R(x,x') &\equiv \rho_{\tau}(x')\pi_R(\overline{x'} \to \overline{x}),\cr
P_{R|m}(x,x') &\equiv \rho_{\tau|m}(x')\pi_R(\overline{x'} \to \overline{x}).
\end{align}
Although the reverse transition functions $\pi_R$ are written as functions of the time-reversed states $\overline x$ and $\overline{x'}$, it is straightforward to apply the time-reversal map on these states to obtain the left-hand side purely as a function of $x$ and $x'$.

It is helpful to think of these reverse joint probabilities in terms of a brand new experiment that starts fresh and runs for time $\tau$. The initial distribution for this experiment is given by the final distribution coming from the forward experiment (with or without updates), and the experiment consists of time-reversing the state, executing the reverse protocol, and then time-reversing the state once more.

Our formalism should be contrasted with the typical formulation of a reverse experiment found in the literature. The initial distribution for the reverse experiment is frequently taken to be the equilibrium distribution for the final choice of control parameters \cite{jarzynski2011equalities}. The present method is more similar to the formalism of Seifert \cite{seifert2008stochastic} in which an arbitrary final distribution, $p_{1}(x_t)$, is considered.

Note that in the definition of $P_{R|m}$, unlike in \eqref{eq:P_F|m} above, the conditioning on $m$ does not affect the transition function $\pi_R$. This is because, from the point of view of the reverse experiment, the measurement happens at the beginning. But $\pi_R$ is a conditional probability which assumes a particular initial state (in this case $\overline{x'}$), and so the measurement $m$ does not provide any additional information that can possibly affect the transition function. Also note the ordering of the arguments as compared with $P_F$ in \eqref{eq-jointdef}: the initial state for the reversed experiment is the second argument for $P_R$, while the initial state for the forward experiment is the first argument in $P_F$. Finally, we record the useful identity
\be\label{eq-jarz1}
\frac{P_{F|m}(x,x')}{P_{R|m}(x,x')} = \frac{P_F(x,x')}{P_R(x,x')},
\ee
assuming both sides are well-defined for the chosen states $x$ and $x'$.

\subsection{Heat Flow}

The Crooks Fluctuation Theorem \cite{Crooks:1999} relates forward and reverse transition functions between equilibrium states to entropy production. It can be thought of as arising via coarse-graining from the ``detailed fluctuation theorem,'' which relates the probabilities of individual forward and backward trajectories to the heat generated along the path through phase space \cite{crooks1998nonequilibrium,jarzynski2011equalities}. Outside the context of equilibrium thermodynamics, this relationship can be thought of as the definition of the ``heat flow'':
\be
Q[x(t)] \equiv \log \frac{\pi_F[x(t)]}{\pi_R[\overline{x}(\tau-t)]}.
\ee
The quantity $Q[x(t)]$ can be equated with the thermodynamic heat (flowing out of the system, in this case) in situations where the latter concept makes sense. (More properly, it is the heat flow in units of the inverse temperature of the heat bath, since $Q[x(t)]$ is dimensionless.) However, $Q[x(t)]$ is a more general quantity than the thermodynamic heat; it is well-defined whenever the transition functions exist, including situations far from equilibrium or without any fixed-temperature heat bath. 

In a similar manner, we can use the coarse-grained transition functions (depending on endpoints rather than the entire path) to define the following useful quantity,
\be\label{eq-Qdef}
\mathcal{Q}(x\to x') \equiv \log \frac{\pi_F(x \to x')}{\pi_R(\overline{x'} \to \overline{x})}.
\ee
This quantity $\mathcal{Q}$, the ``generalized heat flow,'' is intuitively a coarse-grained version of the change in entropy of the environment during the transition $x\to x'$ in the forward experiment, though it is well-defined whenever the appropriate transition functions exist. It is this generalized heat flow that will appear in our versions of the Second Law and the Bayesian Second Law.

\section{Second Laws from Relative Entropy}\label{sec-secondLaw}

All of the information about forward and reversed transition probabilities of the system is contained in the joint forward probability distribution $P_F(x, x')$ and reverse distribution $P_R(x,x')$, defined in \eqref{eq-jointdef} and \eqref{eq:P_R}, respectively. The effects of a Bayesian update on a measurement outcome $m$ are accounted for in the distributions $P_{F|m}(x, x')$ and $P_{R|m}(x, x')$, given in \eqref{eq:P_F|m} and \eqref{eq:P_R}. The most concise statements of the Second Law therefore arise from comparing these distributions.

\subsection{The Ordinary Second Law from Positivity of Relative Entropy}

The relative entropy, also known as the Kullback-Leibler divergence \cite{kullback1951}, is a measure of the distinguishability of two probability distributions: 
\be
D(p\|q) \equiv \int dx\,p(x) \log \frac{p(x)}{q(x)} \geq 0.
\label{eq:DKL}
\ee
In a rough sense, $D(p\|q)$ can be thought of as the amount of information lost by replacing a true distribution $p$ by an assumed distribution $q$. Relative entropy is nonnegative as a consequence of the concavity of the logarithm, and only vanishes when its two arguments are identical. In this sense it is like a distance, but with the key property that it is asymmetric in $p$ and $q$, as both the definition and the intuitive description should make clear.

The relative entropy has been used in previous literature to quantify the information loss due to the stochastic evolution of a system. This has been achieved by analyzing path-space or phase-space distributions at a fixed time \cite{PhysRevLett.98.080602, PhysRevE.73.046105, jarzynski2011equalities}. In a similar manner, we compute the relative entropy of the forward probability distribution with respect to the reverse one. However, we think of $P_F(x, x')$ and $P_R(x,x')$ each as single distributions on the space $\Gamma\times\Gamma$, so that

\be
D(P_F\|P_R) = \int dx dx'\, P_F(x, x') \log \frac{P_F(x,x')}{P_R(x,x')}.
\label{eq:DPFPR}
\ee
Into this we can plug the expressions \eqref{eq-jointdef} and \eqref{eq:P_R} for $P_F$ and $P_R$, as well as the relations \eqref{eq:rho-from-P} between those distributions and the single-time distributions $\rho_0(x)$ and $\rho_\tau(x')$, to obtain
\begin{align}
D(P_F\|P_R) & = \int \rho_0(x) \pi_F(x \to x') \left( \log\frac{\rho_0(x)}{\rho_{\tau}(x')}
                        + \log\frac{\pi_F(x \to x')}{\pi_R(\overline{x'} \to \overline{x})} \right) dx dx' \\
 & = S(\rho_\tau) - S(\rho_0) + \int dxdx'~ P_F(x,x') \mathcal{Q}(x\to x').\label{eq:DPFPRexpanded}
\end{align}
Here $S$ is the usual Gibbs or Shannon entropy, 
\be
S(\rho)\equiv -\int \rho(x) \log{\rho(x)}\, dx, 
\ee
and $\mathcal{Q}$ is the generalized heat flow defined by \eqref{eq-Qdef} above. The first two terms in \eqref{eq:DPFPRexpanded} constitute the change in entropy of the system, while the third term represents an entropy change in the environment averaged over initial and final states. We will introduce the notation $\ev{\cdot}_F$ to denote the average of a quantity with respect to the probability distribution $P_F$, 
\be
  \ev{f}_F \equiv \int dxdx'\, P_F(x, x') f(x, x').
\ee
The positivity of the relative entropy \eqref{eq:DPFPR} is therefore equivalent to
\be\label{eq-osl}
\Delta S + \ev{\mathcal{Q}}_F \geq 0,
\ee
with equality if and only if $P_F = P_R$. This is the simplest form of the Second Law; it says that the change in entropy of the system is bounded from below by (minus) the average of the generalized heat $\mathcal{Q}$ with respect to the forward probability distribution. 

The result \eqref{eq-osl} is an information-theoretical statement; in the general case we should not think of $S$ as a thermodynamic entropy or $\ev{\mathcal{Q}}_F$ as the expectation value of a quantity which can be measured in experiments. To recover the thermodynamic Second Law, we must restrict ourselves to setups in which temperature, heat flow, and thermodynamic entropy are all well-defined. In this case, we can interpret $\ev{\mathcal{Q}}_F$ as the expected amount of coarse-grained heat flow into the environment. ``Coarse-grained" here refers to the difference between the endpoint-dependent $\mathcal{Q}(x\to x')$ and the fully path-dependent $Q[x(t)]$ introduced above. By considering the relative entropy of the forward path-space probability $P_F[x(t)]$ with respect to the reverse one $P_R[x(t)]$, we can recover the ordinary Second Law with the ordinary heat term, obtained from \eqref{eq-osl} by the replacement $\mathcal{Q} \to Q$. We will have more to say about the relationship between these two forms of the ordinary Second Law in the following section.

\subsection{A Refined Second Law from Monotonicity of Relative Entropy}
\label{refinedsecondlaw}

Given any pair of probability distributions $p(x,y)$, $q(x,y)$ on multiple variables, we have
\be
D(p(x,y) \| q(x,y)) \geq D\left(\int dy\,p(x,y)\right|\left| \int dy\,q(x,y)\right).\label{eq:monotonicity}
\ee 
This property is known as the monotonicity of relative entropy.  To build intuition, it is useful to first consider a more general property of the relative entropy:
\be
D(p\|q) \geq D\left(Wp\|Wq\right)\quad \forall W,\label{eq:data processing}
\ee
where $W$ is a probability-conserving (i.e., stochastic) operator. This result follows straightforwardly from the definition of relative entropy and the convexity of the logarithm. In words, it means that performing any probability-conserving operation $W$ on probability distributions $p$ and $q$ can only reduce their relative entropy.  

In information theory, \eqref{eq:data processing} is known as the Data Processing Lemma \cite{Csiszar:1982, cover1991elements, cohen1993majorization}, since it states that processing a signal only decreases its information content. Marginalizing over a variable is one such way of processing (it is probability-conserving by the definition of $p$ and $q$), so marginalization, in particular, cannot increase the relative information. Intuitively, \eqref{eq:monotonicity} says that marginalizing over one variable decreases the amount of information lost when one approximates $p$ with $q$.

Our single-time probability distributions $\rho_t(x)$ can be thought of as marginalized versions of the joint distribution $P_F(x, x')$, following \eqref{eq:rho-from-P}. We can also define a new ``cycled'' distribution by marginalizing $P_R(x,x')$ over $x'$ to obtain
\be
\tilde{\rho}(x) \equiv \int dx' ~P_R(x,x') = \int dx' \rho_\tau(x')\pi_R(\overline{x'}\to \overline{x}).
\label{eq:cycled-distribution}
\ee
This is the probability distribution we find at the conclusion of the reversed experiment, or, in other words, after running through a complete cycle of evolving forward, time-reversing the state, evolving with the reverse protocol, and then time-reversing once more. In the absence of environmental interaction, we expect the cycled distribution to match up with the initial distribution $\rho_0(x)$, since the evolution of an isolated system is completely deterministic.

Applying monotonicity to $P_F$ and $P_R$ by marginalizing over the final state $x'$, we have
\be
D(P_F \| P_R) \geq D(\rho_0\| \tilde{\rho}) \geq 0,
\ee
or simply, using the results of the previous subsection,
\be
\Delta S + \ev{\mathcal{Q}}_F \geq D(\rho_0\| \tilde{\rho}) \geq 0.
\label{eq:refined-second-law}
\ee
This is a stronger form of the ordinary Second Law. It states that the change in entropy is bounded from below by an information-theoretic quantity that characterizes the difference between the initial distribution $\rho_0$ and a cycled distribution $\tilde{\rho}$ that has been evolved forward and backward in time.

In the context of a numerical simulation, it is easier to calculate $D(\rho_0\| \tilde{\rho})$ than $D(P_F \| P_R)$, since the former only depends on knowing the probability distribution of the system at two specified points in time. $D(\rho_0\| \tilde{\rho})$ can readily be calculated by evolving the distribution according to the forward and reverse protocols. This is in contrast with $D(P_F \| P_R)$, the computation of which requires knowledge of joint probability distributions. Obtaining the joint distributions is more difficult, because one must know how each microstate at the given initial time relates to the microstates of the future time. This bound therefore provides an easily-calculable contraint on the full behavior of the system.

Monotonicity of the relative entropy also allows us to succinctly state the relationship between the path-space and endpoint-space formulations of the Second Law. Indeed, the relationship between the probabilities $P_F[x(t)]$ and $P_F(x,x')$ is
\be
P_F(x,x') =  \int_{x(0) = x}^{x(\tau) = x'} \mathcal{D}x(t) \,P_F[x(t)],
\ee
with a similar relationship between the reversed quantities. Monotonicity of relative entropy then implies that
\be
D(P_F[x(t)] \| P_R[x(t)] ) \geq D(P_F(x,x') \| P_R(x,x')).
\ee
Since the changes in entropy are the same, this inequality reduces to the relationship $\ev{Q[x(t)]}_F \geq \ev{\mathcal{Q}(x\to x')}_F $ between the expected heat transfer and the expected coarse-grained heat transfer, which can also be shown directly with a convexity argument. The point here is that the path-space and endpoint-space formulations of the ordinary Second Law (as well as the Bayesian Second Law in the following section) are not independent of each other. Endpoint-space is simply a coarse-grained version of path-space, and the monotonicity of relative entropy tells us how the Second Law behaves with respect to coarse-graining. 

\section{The Bayesian Second Law}\label{sec-BSL}

Now we are ready to include Bayesian updates. It is an obvious extension of the discussion above to consider the relative entropy of the updated joint probabilities $P_{F|m}$ and $P_{R|m}$, which is again non-negative:
\be\label{eq-bsl}
D(P_{F|m}\|P_{R|m}) \geq 0.
\ee
This is the most compact form of the Bayesian Second Law (BSL).

\subsection{Cross-Entropy Formulation of the BSL}

It will be convenient to expand the definition of relative entropy in several different ways. First, we can unpack the relative entropy to facilitate comparison with the ordinary Second Law:
\be\label{eq-updatedRelEntropy}
D(P_{F|m}\|P_{R|m}) = \int dx\, \rho_{0|m}(x)\log \rho_0(x) -  \int dx'\, \rho_{\tau|m}(x')\log \rho_\tau(x') + \ev{\mathcal{Q}}_{F|m}.
\ee
Here we have used the expressions \eqref{eq:P_F|m} and \eqref{eq:P_R} for the joint distributions, as well as the identity \eqref{eq-jarz1}.
We have also extracted the generalized heat term,
\be
\ev{\mathcal{Q}}_{F|m} \equiv \int dxdx'~P_{F|m}(x,x') \log \frac{\pi_F(x\to x')}{\pi_R(\overline{x'}\to \overline{x})},
\ee
which is the expected transfer of generalized heat out of the system during the forward experiment given the final measurement outcome. This is an experimentally measurable quantity in thermodynamic setups: the heat transfer is measured during each trial of the experiment, and $\ev{\mathcal{Q}}_{F|m}$ is the average over the subset of trials for which the measurement outcome was $m$. The remaining two terms are not identifiable with a change in entropy, but we have a couple of options for interpreting them. 

The form of \eqref{eq-updatedRelEntropy} naturally suggests use of the cross entropy between two distributions, defined as 
\be
  H(p,q) = -\int dx\, p(x) \log q(x). 
  \label{cross-entropy}
\ee
(Note that this is not the joint entropy, defined for a joint probability distribution $p(x,y)$ as $-\int dxdy\, p(x,y)\log p(x,y)$.)
Using this definition, the relative entropy between the updated joint distributions \eqref{eq-updatedRelEntropy} may be rewritten in the form,
\be
D(P_{F|m}\|P_{R|m}) = H(\rho_{\tau|m}, \rho_\tau) - H(\rho_{0|m}, \rho_0)   + \ev{\mathcal{Q}}_{F|m}.
\ee
The Bayesian Second Law is then
\be
  \Delta H(\rho_m, \rho) + \ev{\mathcal{Q}}_{F|m} \geq 0.
  \label{cross-entropy-BSL}
\ee
Here, $\Delta$ is the difference in the values of a quantity evaluated at the final time $\tau$ and the initial time $0$.

To get some intuition for how to interpret this form of the BSL, it is useful to recall the information-theoretic meaning of the entropy and cross entropy. Given a probability distribution $p(x)$ over the set of microstates $x$ in a phase space $\Gamma$, we can define the self-information (or Shannon information, or ``surprisal'') associated with each state,
\be
  I_p(x) = \log\frac{1}{p(x)}.
  \label{eq:Self-Info}
\ee
The self-information measures the information we would gain by learning the identity of the specific microstate $x$. If $x$ is highly probable, it's not that surprising to find the system in that state, and we don't learn that much by identifying it; if it's improbable we have learned a great deal.
From this perspective, the entropy $S(p)= \int dx\, p(x)I_p(x)$ is the expectation value, with respect to $p(x)$, of the self-information associated with $p(x)$ itself. It is how much we are likely to learn, on average, by finding out the actual microstate of the system. In a distribution that is highly peaked in some region, the microstate is most likely to be in that region, and we don't learn much by finding it out; such a distribution has a correspondingly low entropy. In a more uniform distribution, we always learn something by finding out the specific microstate, and the distribution has a correspondingly higher entropy.

In contrast, the cross entropy $H(p,q) = \int dx\, p(x)I_q(x)$ is the expectation value with respect to $p(x)$ of the self-information associated with $q(x)$. Typically $p(x)$ is thought of as the ``true'' or ``correct'' distribution, and $q(x)$ as the ``assumed'' or ``wrong'' distribution. We believe that the probability distribution is given by $q(x)$, when it is actually given by $p(x)$. The cross entropy is therefore a measure of how likely we are to be surprised (and therefore learn something) if we were to be told the actual microstate of the system, given that we might not be using the correct probability distribution. The cross entropy is large when the two distributions are peaked, but in different places; that maximizes the chance of having a large actual probability $p(x)$ for a state with a large self-information $I_q(x)$. When the two distributions differ, we are faced with two distinct sources of uncertainty about the true state of the system: the fact that there can be uncertainty in the true distribution, and the fact that we are working with an assumed distribution rather than the true one. Mathematically, this is reflected in the cross entropy being equal to the entropy of the true distribution plus the relative entropy:
\be
  H(p,q) = S(p) + D(p\|q).
\ee
The cross entropy is always greater than the entropy of the true distribution (by positivity of relative entropy), and reduces to the ordinary entropy when the two distributions are the same.

The Bayesian Second Law, then, is the statement that the cross entropy of the updated (``true'') distribution with respect to the original (``wrong'') distribution, plus the generalized heat flow, is larger when evaluated at the end of the experiment than at the beginning. 
In other words, for zero heat transfer, the expected amount of information an observer using the original distribution function would learn by being told the true microstate of the system, conditioned on an observation at the final time, is larger at the final time than at the initial one. 

We note that the quantity $H(\rho_{t|m}, \rho_t)$ only has operational meaning once a measurement has occurred, since performing the Bayesian update to take the measurement into account requires knowledge of the actual measurement outcome. The BSL is a statement about how much an experimenter who knows the measurement outcome would expect someone who didn't know the outcome to learn by being told the microstate of the system. There is therefore not any sense in which one can interpret an increase of $H(\rho_{t|m}, \rho_t)$ with increasing $t$ as an increase in a dynamical quantity. This is in contrast with the dynamical interpretation of the monotonic increase in entropy over time in the ordinary Second Law. It is, in fact, the case that $H(\rho_{t|m}, \rho_t)$ \emph{does} increase with increasing $t$ for zero heat transfer, but this increase can only be calculated retroactively once the measurement has actually been made. Of course, in the case of a trivial measurement that tells us nothing about the system, the BSL manifestly reduces to the ordinary Second Law, since $H(\rho,\rho) = S(\rho)$.

\subsection{Alternate Formulations of the BSL}

Another natural quantity to extract is the total change in entropy after the two-step process of time evolution and Bayesian updating, which we will call $\Delta S_m$:
\be
\Delta S_m \equiv S(\rho_{\tau|m}) - S(\rho_0).
\label{eq:delta-sm}
\ee
This is the actual change in the entropy over the course of the experiment in the mind of the experimenter, who initially believes the distribution is $\rho_0$ (before the experiment begins) and ultimately believes it to be $\rho_{\tau|m}$. In terms of this change in entropy, we have
\be
D(P_{F|m}\|P_{R|m}) = \Delta S_m +\ev{\mathcal{Q}}_{F|m} + D(\rho_{\tau|m}\|\rho_\tau) + \int dx~ (\rho_{0|m}(x) - \rho_0(x))\log \rho_0(x).
\label{eq:D-with-information-gain}
\ee
The second to last term, $D(\rho_{\tau|m}\|\rho_\tau)$, is the relative entropy of the posterior distribution at time $\tau$ with respect to the prior distribution; it can be thought of as the amount of information one gains about the final probability distribution due to the measurement outcome. This is a natural quantity in Bayesian analysis, called simply the \textit{information gain} \cite{lindley1956}; maximizing its expected value (and hence the expected information learned from a measurement) is the goal of Bayesian experimental design \cite{chaloner1995}. Because it measures information gained, it tends to be largest when the measurement outcome $m$ was an unlikely one from the point of view of $\rho_\tau$. The final term exactly vanishes in the special case where the initial probability distribution is constant on its domain, which is an important special case we will consider in more detail below. 

Using \eqref{eq:D-with-information-gain}, the positivity of relative entropy is equivalent to
\be\label{eq-bsl2}
\Delta S_m + \ev{\mathcal{Q}}_{F|m} \geq -D(\rho_{\tau|m}\|\rho_\tau) + \int dx~ (\rho_0(x) - \rho_{0|m}(x))\log \rho_0(x).
\ee
The left-hand side of this in equality is similar to that of the ordinary Second Law, except that the result of the measurement is accounted for. In the event of an unlikely measurement, we would intuitively expect that it should be allowed to be negative. Accordingly, on the right-hand side we find that it is bounded from below by a quantity that can take on negative values. And indeed, the more unlikely the measurement is, the greater $D(\rho_{\tau|m}\|\rho_\tau)$ is, and thus the more the entropy is allowed to decrease.

Finally, we can expand the relative entropy in terms of $S(\rho_{0|m})$ instead of $S(\rho_0)$. That is, we define the change in entropy between the initial and final updated distributions, 
\be
  \Delta S(\rho_m) \equiv S(\rho_{\tau|m}) -S(\rho_{0|m}).
\ee
(Note the distinction between $\Delta S(\rho_m)$ here and $\Delta S_m$ in \eqref{eq:delta-sm}.)
This is the answer to the question, ``Given the final measurement, how much has the entropy of the system changed?''
Then \eqref{eq-bsl2} is equivalent to
\be
  \Delta S(\rho_{m}) + \ev{\mathcal{Q}}_{F|m} \geq D(\rho_{0|m}\|\rho_0)-D(\rho_{\tau|m}\|\rho_\tau).
  \label{eq-bsl3}
\ee
This change of entropy can be contrasted with $S(\rho_{\tau|m}) -S(\rho_{0})$, which is a statement about the change in the experimenter's knowledge of the system before and after the measurement is performed.

The right hand side of \eqref{eq-bsl3} has the interesting property that it is always less than or equal to zero. This can be shown by taking the difference of the relative entropies and expressing it in the form
\be
D(\rho_{0|m}\|\rho_0)-D(\rho_{\tau|m}\|\rho_\tau) = \int dx dx' \frac{\rho_{0}(x) \pi_{F}(x\rightarrow x') P(m|x')}{P(m)} \log \frac{\pi_{F}(x\rightarrow m)}{P(m|x')}.
\ee
We have defined $\pi_{F}(x\rightarrow m) \equiv \int dx' \pi_{F}(x\rightarrow x') P(m|x')$ for convenience. It is only possible to write the difference in this form because the initial and final distributions are related by evolution \eqref{updated-final-pi}. Using the concavity of the logarithm, it can then be shown that this quantity is non-positive.

One final  point of interest in regards to \eqref{eq-bsl3} is its average with respect to measurement outcomes. The inequality is predicated on a specific measurement outcome, $m$; averaging with respect to the probability of obtaining a given measurement, we find
\be
\left\langle \Delta S(\rho_{m}) \right\rangle + \left\langle \mathcal{Q} \right\rangle \geq I(X_0;M)-I(X_\tau;M)
\label{eq-bsl-ave}
\ee
where $I(X_t;M)$ is the mutual information between the microstate of the system at time $t$ and the measurement outcome. Here the mutual information can be expressed as the relative entropy of a joint probability distribution to the product of its marginal distributions, $I(x;m)=D(\rho(x,m)\|\rho(x)\rho(m))$.

Inequalities similar to \eqref{eq-bsl-ave} can be found in the existing literature for nonequilibrium feedback-control, though they are usually written in terms of work and free energy instead of entropy \cite{PhysRevLett.100.080403, PhysRevLett.104.090602, sagawa2012nonequilibrium, PhysRevE.82.061120, PhysRevE.82.031129}. The novelty of \eqref{eq-bsl-ave} stems from the fact that no explicit feedback-control is performed after the measurement and the presence of the term $I(X_0;M)$. This term is the mutual information between the initial microstate of the system and the measurement outcome and arises because Bayesian updating is performed at the initial time as well as the final time. Due to this updating, the lower bound on the entropy production is greater than one would naively suspect without updating the initial state.

\subsection{A Refined BSL from Monotonicity of Relative Entropy}\label{sec-refinedBSL}

So far we have rewritten the relative entropy of the forward and reverse distributions \eqref{eq-bsl} in various ways, but there is a refined version of the BSL that we can formulate using monotonicity of relative entropy, analogous to the refined version of the ordinary Second Law we derived in Section~\ref{refinedsecondlaw}. Following the definition of the cycled distribution $\tilde{\rho}$ in \eqref{eq:cycled-distribution}, we can define an updated cycled distribution by marginalizing the updated reverse distribution over initial states,
\be
\label{eq:updated-cycled-distribution}
\tilde{\rho}_m(x) \equiv \int dx' ~P_{R|m}(x,x') = \int dx' \rho_{\tau|m}(x')\pi_R(\overline{x'}\to \overline{x}).
\ee
The monotonicity of relative entropy then implies that
\be
D(P_{F|m} \| P_{R|m}) \geq D(\rho_{0|m}\| \tilde{\rho}_m).
\label{eq:refined-bsl}
\ee
This is the refined Bayesian Second Law of Thermodynamics in its most compact form, analogous to the refined Second Law \eqref{eq:refined-second-law}. 

Expanding the definitions as above, the refined BSL can be written as
\be
  \Delta H(\rho_m, \rho) + \ev{\mathcal{Q}}_{F|m} \geq D(\rho_{0|m}\| \tilde{\rho}_m),
  \label{eq:refined-BSL2}
\ee
or equivalently as
\be
\label{eq:expanded-refined-bsl}
\Delta S_m + \ev{\mathcal{Q}}_{F|m} \geq D(\rho_{0|m}\| \tilde{\rho}_m)-D(\rho_{\tau|m}\|\rho_\tau) + \int dx~ (\rho_0(x) - \rho_{0|m}(x))\log \rho_0(x).
\ee
From the form of \eqref{eq:refined-BSL2}, we see that the change in the cross entropy obeys a tighter bound than simple positivity, as long as the cycled distribution deviates from the original distribution (which it will if the evolution is irreversible).

Other versions of the Second Law can be obtained from the relative entropy by inserting different combinations of $P_{F|m}$, $P_{R|m}$, $P_F$, and $P_R$. We have chosen to highlight $D(P_F\|P_R)$ and $D(P_{F|m}\|P_{R|m})$ because these are the combinations which we can expect to vanish in the case of perfect reversibility, and thus characterize the time-asymmetry of the dynamics. Other possibilities, like $D(P_{F|m}\|P_R)$, are always nonzero as long as information is gained from the measurement. 

\section{Bayesian Jarzynski Equalities}
\label{sec:Bayesian-Jarzynski}

The Jarzynski equality \cite{Jarzynski:1997} relates the expectation value of the work done on a system along paths connecting two equilibrium states (the paths themselves can involve non-equilibrium processes). Consider two equilibrium states $A$ and $B$, with Helmholtz free energies $F_A$ and $F_B$, and define $\Delta F \equiv F_B-F_A$. If the work done as the system evolves from $A$ to $B$ is denoted by $W$, the Jarzynski equality states
\be
  \ev{e^{-W}} = e^{-\Delta F}.
  \label{eq:Jarzynski}
\ee
(As usual we are setting Boltzmann's constant $k_B$ and the inverse temperature $\beta$ equal to unity.)
This equality represents a non-equilibrium relationship between the set of all paths between two states and their respective free energies.  It can be derived from the Crooks Fluctuation Theorem \cite{Crooks:1999}, which can be written as
\be
  \frac{P(A\rightarrow B)}{P(B\rightarrow A)} = e^{W-\Delta F},
  \label{eq:crooks-ft}
\ee
where $P(A\rightarrow B)$ and $P(B\rightarrow A)$ are respectively the forward and reverse probabilities between these equilbrium states. In turn, \eqref{eq:Jarzynski} immediately implies the Second Law via Jensen's inequality, $\ev{e^x} \geq e^{\ev{x}}$, given that $\Delta S = W - \Delta F$ between equilibrium states. In this section we show how to derive a number of equalities involving expectation values of quotients of forward and reverse probabilities, with and without Bayesian updates. For simplicity we will refer to such relations as ``Jarzynski equalities."

Recall the simple identity \eqref{eq-jarz1}: 
\be
\frac{P_R(x,x')}{P_F(x,x')} = \frac{P_{R|m}(x,x')}{P_{F|m}(x,x')} = \frac{ \rho_\tau(x') }{\rho_0(x)}e^{-\mathcal{Q}(x\to x')},
\label{eq:simple-identity}
\ee
which we have made use of in previous sections.
We can obtain a Jarzynski equality by computing the expectation value of this ratio with respect to $P_F$ (or $P_{F|m}$). Naively, one would multiply by $P_F$ and find $P_F P_R/P_F = P_R$, but we need to keep track of the domain of integration: we are only interested in points where $P_F \neq 0$ ($P_{F|m} \neq 0$) when computing an average with respect to $P_F$ ($P_{F|m}$). So we have, for instance,
\be\label{eq-jarz2}
\ev{\frac{P_R}{P_F}}_F = \int_{P_F\neq 0}dxdx'~P_R(x,x').
\ee
This integral will be equal to one unless there is a set of zero $P_F$-measure with nonzero $P_R$-measure. On such a set, the ratio $P_R/P_F$ diverges. Generically this will include all points where $\rho_0(x)$ vanishes, unless $\mathcal{Q}$ happens to diverge for some choices of $x'$ ({\it e.g.}, one reason for $P_R$ to vanish is that certain transitions are strictly irreversible). Note that if $\rho_0(x)$ is nowhere zero and $\mathcal{Q}$ does not ever diverge (as in physically relevant situations), then this integral is equal to one. This is true no matter how small $\rho_0(x)$ is or how large $\mathcal{Q}$, as long as they are nonzero and finite everywhere, respectively. For this reason, \eqref{eq-jarz2} generically is equal to one.

The same reasoning holds for the updated probabilities:
\be
\ev{\frac{P_{R|m}}{P_{F|m}}}_{F|m} = \int_{P_{F|m}\neq 0}dxdx'~P_{R|m}(x,x').
\ee
Since the ratio $P_{R|m}/P_{F|m}$ is identical to the ratio $P_R/P_F$, the condition for this integral to equal one is the same as the previous integral, which means it is generically so.

To summarize, we have constants $a,b_m$ such that
\be
\ev{\frac{P_R}{P_F}}_F =a \leq 1,~~~\ev{\frac{P_{R|m}}{P_{F|m}}}_{F|m}=b_m  \leq 1.
\label{eq:bje}
\ee
By perturbing the initial state by an arbitrarily small amount, we can make $P_R/P_F$ finite everywhere (excluding divergences in $\mathcal{Q}$), and so $a\neq 1$ and $b_m\neq 1$ are in some sense unstable. As with the usual Jarzynski equality, we can use Jensen's inequality on each of these to extract a Second Law:
\begin{align}
D(P_F\|P_R)&\geq - \log a \geq 0\\
D(P_{F|m}\|P_{R|m}) &\geq -\log b_m \geq 0. 
\end{align}
Thus these Jarzynski equalities contain within them the positivity of relative entropy.

There are also Jarzynski equalities corresponding to the monotonicity inequalities. Consider
\be\label{eq-jarz3}
\ev{\frac{P_R}{P_F} \frac{\rho_0}{\tilde\rho}}_F =  \int_{P_F\neq 0}dxdx'~ \frac{P_R(x,x')}{\int dy'~P_R(x,y')} \rho_0(x) \leq  1.
\ee
Applying Jensen's inequality reproduces the monotonicity result:
\be
D(P_F\|P_R) \geq D(\rho_0\|\tilde\rho).
\ee
The refined Bayesian Second Law follows similarly from the Jarzynski equality,
\be
\ev{\frac{P_{R|m}}{P_{F|m}} \frac{\rho_{0|m}}{\tilde\rho_m}}_{F|m} = c_{m} \leq 1.
\ee

While we have derived a series of Jarzynski equalities, it is not apparent that these share the mathematical form of the original Jarzynski equality, \eqref{eq:Jarzynski}; specifically, we would like to make contact with an exponential average of thermodynamic quantities. Making use of the identity \eqref{eq:simple-identity}, we may write

\be
\ev{\frac{P_R}{P_F}}_F =  \ev{e^{\log\rho_{\tau}(x') - \log\rho_{0}(x) -\beta \mathcal{Q}(x \rightarrow x') }}_F = a.
\label{eq:Jarzynski-like}
\ee

A similar equality may be derived making use of the thermodynamic heat, $Q$, instead of the coarse-grained heat, $\mathcal{Q}$; however, in doing so, one must introduce an average over paths. We note that \eqref{eq:Jarzynski-like} takes the form of an exponential of a difference in self-informations, \eqref{eq:Self-Info}, and heat transfer. For initial and final macrostates where the self-information of each microstate is equal to the macrostate's entropy, this reduces to Jarzynski's original equality. As such, \eqref{eq:Jarzynski-like} is just a restatement of the Jarzynski equality generalized for non-equilibrium states.

In a similar manner, we also find:

\be
\ev{e^{\log\rho_{\tau}(x') - \log\rho_{0}(x) -\beta \mathcal{Q}(x \rightarrow x') }}_{F|m} = b_{m},
\label{eq:Jarzynski-updated}
\ee

\be
\ev{e^{\log\rho_{\tau}(x') - \log\rho_{0}(x) -\beta \mathcal{Q}(x \rightarrow x') + \log[\rho_{0|m}(x)/\tilde\rho_m(x)]}}_{F|m} = c_{m}.
\label{eq:Jarzynski-cycled}
\ee
We see that \eqref{eq:Jarzynski-updated} generalizes the Jarzynski equality to include Bayesian updating while \eqref{eq:Jarzynski-cycled} also includes the information loss from the stochastic evolution of the system. Importantly, \eqref{eq:Jarzynski-updated} and \eqref{eq:Jarzynski-cycled} hold independently for each possible measurement outcome. Essentially, if we partition a large set of experimental trials based on measurement outcomes, each subset obeys its own Jarzynski equality, \eqref{eq:Jarzynski-updated}. However, if we consider all experimental trials together the Jarzynski equality \eqref{eq:Jarzynski-like} holds. This leads us to the relation

\be
\int dm ~P(m) b_{m} = a.
\ee

\section{Applications}

As a way to build some intuition for the Bayesian point of view we have been discussing, we will go through a few simple examples and special cases.

\subsection{Special Cases}\label{sec-specialcases}

\paragraph{Perfect Complete Measurement.}
If a measurement does not yield any no new information, then the updated probabilities are identical to the prior probabilities and the Bayesian Second Law reduces to the ordinary Second Law. On the other hand, consider a measuring device that is able to tell us with certainty what the exact microstate of the system is at the time of measurement. 
The outcome $m$ of the experiment is then a single point in phase space. If we employ such a device, we have the following simplified expressions:
\begin{align}
\rho_{0|m}(x) &= \frac{\rho_0(x)\pi_F(x\to m)}{\rho_\tau(m)}, \\
\rho_{\tau|m}(x') &= \delta(x'-m),\\
\pi_{F|m}(x \to x') &= \delta(x'-m)\theta(\pi_F(x\to m)),\\
\tilde \rho_m(x) &= \pi_R(\overline{m} \to \overline{x}).
\end{align}
Using these simplifications, we find
\be
D(P_{F|m}\|P_{R|m}) = D(\rho_{0|m}\|\tilde{\rho}_m),
\ee
so the refined Bayesian Second Law is always saturated. This is because marginalization of the joint distribution over the final endpoint results in no loss of information: we are still conditioning on the measurement outcome $m$, which tells us the final endpoint.

\paragraph{The Boltzmann Second Law of Thermodynamics.}
In the Boltzmann formulation of the Second Law, phase space is partitioned into a set of macrostates. Each microstate is assigned to a macrostate; the entropy of a microstate $x$ is defined as the entropy of its associated macrostate $\Sigma(x)$, which is the logarithm of the macrostate's phase space volume $|\Sigma|$. We can reproduce this formulation as a special case of the Bayesian measurement formalism: the measuring device determines which macrostate the microstate belongs to with absolute certainty. If the measurement outcome $m$ indicates that the system is in some particular macrostate (but doesn't include any additional information), we have
\be
P(m | x) = \mathbbm{1}_m(x)  \equiv \begin{cases}
1~{\rm if}~x\in m,\\
0~{\rm if}~x\not\in m.
\end{cases}
\ee
We also choose our initial distribution to be uniform over an initial macrostate $\Sigma_0$:
\be
\rho_0(x) =  \frac{1}{|\Sigma_0|}\mathbbm{1}_{\Sigma_0}(x).
\ee
Then we have the identities
\begin{align}
\ev{-  \log  \rho_{0}(x)}_{F|m}  &= \log |\Sigma_0| = S(\rho_0),\\
\ev{-  \log  \rho_{\tau}(x)}_{F|m}  &= -\int dx~\rho_{\tau | m}(x)\log\rho_{\tau |m}(x)+\int dx~\rho_{\tau | m}(x)\log\frac{\rho_{\tau |m}(x)}{\rho_{\tau}(x)} \\
&= S(\rho_{\tau|m}) + D(\rho_{\tau | m}\| \rho_{\tau}).
\end{align}
Then the refined Bayesian Second Law \eqref{eq:expanded-refined-bsl} simplifies to 
\be\label{eq-boltz}
\Delta S_m + \ev{\mathcal{Q}}_{F|m} \geq D(\rho_{0|m} \| \tilde \rho_m) - D(\rho_{\tau|m} \| \rho_\tau ).
\ee

The left-hand side of this inequality is not quite the same as in the Boltzmann formulation, because $S(\rho_{\tau|m})$ is not the entropy associated with any of the previously established macrostates. But we do have the inequality $S(\rho_{\tau|m}) \leq \log |m|$, which {\em is} the entropy of the final macrostate. So the left-hand side of \eqref{eq-boltz} can be replaced by the usual left-hand side of the Boltzmann Second Law while maintaining the inequality.\footnote{And, as we have discussed previously, the coarse-grained $\mathcal{Q}$ can be replaced by the path-space $Q$ as well.} 

The right-hand side of the Boltzmann Second Law is zero, while in \eqref{eq-boltz} we have the difference of two positive terms. The Boltzmann Second Law can be violated by rare fluctuations, and here we are able to characterize such fluctuations by the fact that they render the right-hand side of our inequality negative. We can also give an explicit formula for the term $D(\rho_{\tau|m} \| \rho_\tau )$ that comes in with a minus sign:
\be
D(\rho_{\tau|m} \| \rho_\tau ) = -\log \int_m dx'~ \rho_\tau(x')= -\log P(m) = I_m,
\ee
where $I_m$ is the self-information associated with the measurement outcome $m$.
When the observed measurement is very surprising, the entropy change has the opportunity to become negative. This gives quantitative meaning to the idea that we gain information when we observe rare fluctuations to lower-entropy states. In particular, the entropy change may be negative if the information gain from the measurement is greater than the information loss due to irreversible dynamics.

\subsection{Diffusion of a Gaussian in $n$ Dimensions.}
As our final analytic example, we consider a dynamical model that can be solved analytically. Let the configuration space be $\mathbb{R}^n$, and suppose the time evolution of the probability density is diffusive. That is, 
\be
\rho_\tau(x') = \int d^nx \frac{1}{(2\pi D\tau)^{n/2}} e^{-\frac{|x-x'|^2}{2D \tau}} \rho_0(x).
\ee
Then we can identify the transition function with the heat kernel:
\be
\pi_F(x \to x') = \frac{1}{(2\pi D\tau)^{n/2}} e^{-\frac{|x-x'|^2}{2D \tau}} .
\ee
We will assume for simplicity that the diffusion is unaffected by time reversal, so that $\pi_F = \pi_R \equiv \pi$, and that the states $x$ are also unaffected by time reversal. (Alternatively, we can assume that time-reversal is some sort of reflection in $x$. The distributions we consider will be spherically symmetric, and hence invariant under such reflections.) Note that since $\pi(x\to x') = \pi(x'\to x)$, this implies $\mathcal{Q} = 0$. We will analyze the system without including measurement, again for simplicity, and we will also assume that the initial density profile is Gaussian with initial width $\sigma$. Diffusion causes the Gaussian to spread:
\be
\rho_\tau(x) = \frac{1}{(2\pi (\sigma + D\tau))^{n/2}}e^{-\frac{x^2}{2(\sigma+D \tau)}} .
\ee
We can also calculate the entropy as a function of time:
\begin{align}
S(\tau) &= \int d^nx \frac{1}{(2\pi( \sigma + D\tau))^{n/2}}e^{-\frac{x^2}{2(\sigma+D \tau)}}\left[\frac{x^2}{2(\sigma+D \tau)} + \frac{n}{2} \log (2\pi(\sigma + D\tau))\right]\\
&=\frac{n}{2} \log(\sigma + D\tau)+ \frac{n}{2}\log 2\pi e.
\end{align}
Therefore we have $\Delta S = \frac{n}{2} \log(1+ \frac{D\tau}{\sigma})$. The relative entropy $D(\rho_0\|\tilde{\rho})$ is also easy to calculate, since in this case $\tilde \rho = \rho_{2\tau}$:
\be
D(\rho_0 \| \tilde{\rho}) =\frac{n}{2}\left[ \log \left(1+\frac{2D\tau}{\sigma}\right)- \frac{2D\tau}{\sigma+2D\tau}\right].
\ee
The refined Second Law from monotonicity of the relative entropy says that $\Delta S \geq D(\rho_0 \| \tilde{\rho}) $. Let us see how strong this is compared to $\Delta S \geq0$. For small $\tau$, we have $ D(\rho_0 \| \tilde{\rho}) \approx n(D\tau/\sigma)^2$, as compared to $\Delta S \approx n D\tau/2\sigma$. So the bound from monotonicity is subleading in $\tau$, so perhaps not so important. For large $\tau$, though, we have $ D(\rho_0 \| \tilde{\rho})  \approx \frac{n}{2}\left[\log\frac{D\tau}{\sigma} -\log\frac{e}{2}\right]$, as compared to $\Delta S \approx  \frac{n}{2}\log\frac{D\tau}{\sigma}$. Now the bound is fairly tight, with the relative entropy matching the leading behavior of $\Delta S$.

\subsection{Randomly Driven Harmonic Oscillator}
\label{sec:Oscillator}

As a slightly more detailed -- and potentially experimentally realizable -- example to which we can apply the Bayesian Second Law, we consider the harmonic oscillator. Imagine a single, massive particle confined to a one-dimensional harmonic potential, with spring constant and potential minimum treated as time-dependent control parameters, coupled to a heat bath which generates dissipative and fluctuating forces. 
Such a system may be described by the Fokker-Planck equation,
\begin{align}
\label{eq:FokkerPlanck}
\frac{\partial\rho(x,p,t)}{\partial t} = 
\frac{2}{\tau_{*}}\rho(x,p,t)
&+\left(k(t)\left[x-z\left(t\right)\right]
+\frac{2}{\tau_{*}}p\right)\frac{\partial\rho(x,p,t)}{\partial p}  \cr
&- \frac{p}{M}\frac{\partial\rho(x,p,t)}{\partial x}+\frac{2M}{\beta\tau_{*}}\frac{\partial^{2}\rho(x,p,t)}{\partial p^{2}}.
\end{align}
Here we have defined $\tau_{*}$ to be the dissipation time-scale, $k( t )$ to be the spring constant, $z( t )$ to be the location of the potential's minimum, $M$ to be the mass of the oscillator, and $\beta$ to be the inverse temperature of the heat bath. For simplicity, we choose to work in units natural for this system by taking $\beta=1$, $M=1$, and $k(t=0)=1$. We also choose $\tau_{*}=1$, so that we are in the interesting regime where the dissipation and oscillation time scales are comparable.

We assume that the experimenter is only capable of measuring the position of the particle and not its momentum. For a microstate with position $x$, we assume that $P(m|x)$ is given by a Gaussian distribution in $m$ centered at $x$ with standard deviation $\sigma=0.2$. This means that the experimenter is likely to find a measured value $m$ within a range $\pm0.2$ of the true position $x$. This measuring device is therefore quite sensitive when compared to the typical size of thermal fluctuations, which is of order unity.

There is no analytical solution to \eqref{eq:FokkerPlanck} in the regime of interest, so the system must be modeled numerically. This can be done by discretizing phase space on a lattice and using finite-difference methods to evolve the discrete probability distribution. We have performed this process using the finite element solver package FiPy \cite{FiPy:2009} for the Python programming language. To elucidate different aspects of the BSL, we consider three different simulated experiments. The phase space evolution of these experiments is shown in Figures \ref{fig:evo-static1} - \ref{fig:evo-drag}, found in Appendix \ref{app:Oscillator}, while the thermodynamic quantities calculated are tabulated in Table \ref{tab:SimResults}. The source code which was used to carry out these simulations and animations of the evolution are also available.\footnote{See: \url{http://preposterousuniverse.com/science/BSL/} }

\begin{table}[t]
\begin{center} \begin{tabular}{  l  r  r  r } \hline
  & Figure \ref{fig:evo-static1} & Figure \ref{fig:evo-static2} & Figure \ref{fig:evo-drag} \\ \hline 
$S(\rho_{0})$ & $2.84$ & $0.31$ & $0.31$ \\
$S(\rho_{\tau})$ & $2.91$ & $2.93$ & $2.96$ \\
$\Delta S$ & $0.07$ & $2.61$ & $2.65$ \\
$\left\langle \mathcal{Q} \right\rangle_F$ & $-0.04$ & $5.99$ & $7.99$ \\
$\Delta S+\left\langle \mathcal{Q} \right\rangle_F$ & $0.02$ & $8.61$ & $10.64$ \\
$D(\rho_{0}\| \tilde{\rho})$ & $0.01$ & $7.68$ & $10.64$ \\ \hline
$S(\rho_{0|m})$ & $2.47$ & $-0.43$ & $0.31$ \\
$S(\rho_{\tau|m})$ & $1.23$ & $1.12$ & $1.23$ \\
$\Delta S_m$ & $-1.61$ & $0.81$ & $0.92$ \\
$D(\rho_{0|m}\|\rho_{0})$ & $1.01$ & $0.70$ & $<0.01$ \\
$D(\rho_{\tau|m}\|\rho_{\tau})$ & $2.71$ & $1.37$ & $1.24$ \\ \hline 
$H(\rho_{0|m},\rho_{0})$ & $3.48$ & $0.26$ & $0.31$ \\
$H(\rho_{\tau|m},\rho_{\tau})$ & $3.94$ & $2.49$ & $2.47$ \\
$\Delta H$ & $0.46$ & $2.23$ & $2.16$ \\
$\left\langle \mathcal{Q} \right\rangle_{F|m}$ & $-0.40$ & $6.14$ & $8.47$ \\
$\Delta H+\left\langle \mathcal{Q} \right\rangle_{F|m}$ & $0.06$ & $8.36$ & $10.64$ \\
$D(\rho_{0|m}\|\tilde{\rho}_{m})$ & $0.04$ & $7.65$ & $10.63$ \\ \hline 
LHS of Eqn \ref{eq:expanded-refined-bsl} & $-2.01$ & $6.94$ & $9.39$ \\
RHS of Eqn \ref{eq:expanded-refined-bsl} & $-2.03$ & $6.235$ & $9.39$ \\
$\left|\frac{\textrm{LHS}-\textrm{RHS}}{\textrm{LHS}}\right|$ & $<0.01$ & $0.10$ & $<0.01$ \\ \hline 
$\ev{\frac{P_R}{P_F}}_F$ & $1.00$ & $1.00$ & $1.00$ \\
$\ev{\frac{P_R}{P_F} \frac{\rho_0}{\tilde\rho}}_F$ & $1.00$ & $1.00$ & $1.00$ \\
$\ev{\frac{P_{R|m}}{P_{F|m}}}_{F|m}$ & $1.00$ & $1.00$ & $1.00$ \\
$\ev{\frac{P_{R|m}}{P_{F|m}} \frac{\rho_{0|m}}{\tilde\rho_m}}_{F|m}$ & $1.00$ & $1.00$ & $1.00$ \\ \hline 
\end{tabular} \end{center}
\caption{List of thermodynamic properties calculated for three numerically simulated experiments.}
\label{tab:SimResults}
\end{table}

We first consider the simple experiment shown in Figure \ref{fig:evo-static1}. The system begins in thermal equilibrium, Figure \ref{fig:initDist-static1}. The experiment is carried out under a ``trivial'' protocol, where the experimenter fixes $k(t)=1$ and $z(t)=0$. Under this protocol, the system is allowed to evolve from $t=0$ to $t=1$ before a measurement is performed. As seen in Figure~\ref{fig:finDist-static1}, the thermal distribution is nearly unchanged by this evolution. (Due to finite-size effects, the thermal distribution is not perfectly stationary.) At the end of the experiment, a measurement of the position is made and we assume that the unlikely fluctuation $m=2$ is observed. The experimenter can then use this information to perform a Bayesian update on both the initial and final distributions as shown in Figures \ref{fig:updatedInitDist-static1} and \ref{fig:updatedFinDist-static1}. To evaluate the irreversibility of this experiment, the experimenter must also examine the time-reversed process. The updated cycled distribution which results from evolving under the time-reversed protocol is shown in Figure \ref{fig:updatedCycledDist-static1}.

While this experiment and its protocol are fairly simple, they illustrate several key features of the Bayesian Second Law. Before the final measurement is performed, the experimenter would state that $\Delta S=0.07$. After performing the measurement, this becomes $\Delta S_{m}=-1.61$ with a heat transfer of $\left\langle\mathcal{Q}\right\rangle_{F|m} =-0.40$. Naively using these updated quantities in \eqref{eq-osl} leads to an apparent violation of the usual Second Law of Thermodynamics. However, this is remedied when one properly takes into account the information gained as a result of the measurement. A more careful analysis then shows $\Delta H = 0.46$ and $D(\rho_{0|m}|\tilde{\rho}_{m}) = 0.04$. As such, we see that \eqref{eq:refined-BSL2} is satisfied and that the inequality is very tight. 

We will now consider the same (trivial) protocol with a different initial distribution. The experimenter knows the initial position of the oscillator and the magnitude, but not the direction, of its initial momentum with a high degree of certainty. As such, there are two regions of phase space the experimenter believes the system could be in. The initial distribution is shown in Figure \ref{fig:initDist-static2}. The system is then allowed to evolve until $t=0.5$ as shown in Figure \ref{fig:finDist-static2}. At the end of the experiment, the position of the oscillator is measured to be $m=2$. The impact of this measurement can be seen in Figures \ref{fig:updatedInitDist-static2} and \ref{fig:updatedFinDist-static2}.

Due to the outcome of the measurement, the experimenter is nearly certain that the oscillator had positive initial momentum. One therefore expects this information gain to be roughly one bit and this is confirmed by $D( \rho_{0|m} \| \rho_{0})=0.70 \approx \log 2$. Despite this sizable information gain for the initial distribution, we note that the information gain for the final distribution is even greater with $D( \rho_{\tau|m} \| \rho_{\tau} )=1.37$. This is expected because, regardless of the measurement outcome, the experimenter will always gain at least as much information about the final distribution than the initial when performing a measurement. Evaluating the remaining terms, see Table \ref{tab:SimResults}, we once again find that the BSL is satisfied.

Lastly, consider an experiment that starts with the same initial state but uses a non-trivial protocol where the potential is ``dragged''. The experimenter keeps $k(t)=1$ fixed but varies $z(t)$. For times between $t=0$ and $t=1$, the experimenter rapidly drags the system according to $z(t\leq 1)=2 t$. After this rapid dragging motion, the experimenter keeps $z(t> 1)=2$ and allows the system to approach equilibrium until a measurement performed at $t=5$. Importantly, this gives the system a significant amount of time to reach its new equilibrium distribution before the measurement is performed. The experimenter then measures the oscillator's position and finds it to be centered in the new potential ($m=2$). The evolution of this system is shown in Figure ~\ref{fig:evo-drag}.

Due to the change in protocol, the experimenter gains an appreciable amount of information about the final distribution of the system, but negligible information about the initial distribution. Specifically, we find that $D(\rho_{\tau|m}\| \rho_{\tau}) = 1.24$, while $D(\rho_{0|m} \| \rho_{0} )<0.01$. This is because the system is given time to fully thermalize before the measurement, so any information about the initial state is lost by the time the measurement is performed. Also of interest is the difference between the forward and reverse protocol. As shown in Figures ~\ref{fig:initDist-drag} and \ref{fig:finDist-drag}, the forward protocol results in most distributions reaching the new thermal equilibrium. However, the same is not true of the reverse protocol: the distributions in Figures \ref{fig:revProFinDist-drag} and \ref{fig:conjFinDist-drag} are not near equilibrium. This is due to the asymmetry between the forward and reverse protocols.

We also calculated the quantities appearing in the Bayesian Jarzynski equalities derived in Section \ref{sec:Bayesian-Jarzynski}; they appear in Table \ref{tab:SimResults}. We find that for all three experimental protocols considered, these are well defined and equal to unity.

\section{Discussion}

We have shown how to include explicit Bayesian updates due to measurement outcomes into the evolution of probability distributions obeying stochastic equations of motion, and derived extensions of the Second Law of Thermodynamics that incorporate such updates. Our main result is the Bayesian Second Law, which can be written in various equivalent forms \eqref{eq-bsl}, \eqref{cross-entropy-BSL}, \eqref{eq-bsl2}, \eqref{eq-bsl3}:
\begin{align}
  D(P_{F|m}\|P_{R|m}) &\geq 0, \\
  \Delta H(\rho_m, \rho) + \ev{\mathcal{Q}}_{F|m} & \geq 0, \label{eq-cross-entropy-BSL_7}\\
  \Delta S_m + \ev{\mathcal{Q}}_{F|m} &\geq -D(\rho_{\tau|m}\|\rho_\tau) + \int dx~ (\rho_0(x) - \rho_{0|m}(x))\log \rho_0(x),\label{eq-bsl2_7}\\
   \Delta S(\rho_{m}) + \ev{\mathcal{Q}}_{F|m} &\geq D(\rho_{0|m}\|\rho_0)-D(\rho_{\tau|m}\|\rho_\tau)\label{eq-bsl3_7}.
\end{align}
We also used monotonicity of the relative entropy to derive refined versions of the ordinary Second Law and the BSL, \eqref{eq:refined-second-law} and \eqref{eq:refined-BSL2}:
\begin{align}
\Delta S + \ev{\mathcal{Q}}_F &\geq D(\rho_0\| \tilde{\rho}) \geq 0,\\
 \Delta H(\rho_m, \rho) + \ev{\mathcal{Q}}_{F|m} &\geq D(\rho_{0|m}\| \tilde{\rho}_m)\geq 0.
\end{align}
Finally, we applied similar reasoning to obtain Bayesian versions of the Jarzynski equality, such as \eqref{eq:bje}:
\be
\ev{\frac{P_{R|m}}{P_{F|m}}}_{F|m}=b_m  \leq 1.
\ee
In the remainder of this section we briefly discuss some implications of these results.

\paragraph{Downward fluctuations in entropy.} As mentioned in the Introduction, there is a tension between a Gibbs/Shannon information-theoretic understanding of entropy and the informal idea that there are rare fluctuations in which entropy decreases. The latter phenomenon is readily accommodated by a Boltzmannian definition of entropy using coarse-graining into microstates, but it is often more convenient to work with distribution functions $\rho(x)$ on phase space, in terms of which the entropy of a system with zero heat flow will either increase or remain constant.

The BSL resolves this tension. The post-measurement entropy of the updated distribution $\rho_{\tau|m}$ can be less than the original starting entropy $\rho_{0}$, as the right-hand side of \eqref{eq-bsl2_7} can be negative. On the rare occasions when that happens, there is still a lower bound on their difference. From the information-theoretic perspective, downward fluctuations in entropy at zero heat flow are necessarily associated with measurements. 

This perspective is also clear from the refined Bayesian version of the Boltzmann Second Law \eqref{eq-boltz}, in which the right-hand side can be of either sign. We can see that downward fluctuations in entropy at zero heat flow occur when the amount of information gained by the experimenter exceeds the amount of information lost due to irreversible dynamics.  

The usefulness of the BSL is not restricted to situations in which literal observers are making measurements of the system. We might be interested in fluctuating biological or nanoscale systems in which a particular process of interest necessarily involves a downward fluctuation in entropy. In such cases, even if there are no observers around to witness the fluctuation, we may still be interested in conditioning on histories in which such fluctuations occur, and asking questions about the evolution of entropy along the way. The BSL can be of use whenever we care about evolution conditioned on certain measurement outcomes.

\paragraph{The Bayesian arrow of time.} Shalizi \cite{shalizi2004backwards} has previously considered the evolution of conservative systems with Bayesian updates. For a closed, reversible system, the Shannon entropy remains constant over time, as the distribution evolves in accordance with Liouville's Theorem.
If we occasionally observe the system and use Bayes's rule to update the distribution, our measurements will typically cause the entropy to  decrease, because conditioning reduces entropy when averaged over measurement outcomes, $\langle S(\rho_m)\rangle_m \leq S(\rho)$. At face value, one might wonder about an apparent conflict between this fact and the traditional understanding of the arrow of time, which is based on entropy increasing over time.
This should be a minor effect in realistic situations, where systems are typically open and ordinary entropy increase is likely to swamp any decrease due to conditioning, but it seems like a puzzling matter of principle.

Our analysis suggests a different way of addressing such situations: upon making a measurement, we can update not only the current distribution function, but the distribution function at all previous times as well. As indicated by \eqref{eq-bsl3_7}, the entropy of the updated distribution can decrease even at zero heat transfer. We have identified, however, a different quantity, the cross entropy $H(\rho_m,\rho)$ of the updated distribution with respect to the unupdated one, which has the desired property of never decreasing \eqref{eq-cross-entropy-BSL_7}. For a closed system, both the updated entropy and the cross entropy will remain constant; for open systems the cross entropy will increase.  It is possible to learn about a system by making measurements, but we will always know as much or more about systems in the past than we do about them in the present.

\paragraph{Statistical physics of self-replication.} The application of statistical mechanics to the physics of self-replicating biological systems by England  \cite{england2013statistical} was one of the inspirations for this work. England considers the evolution of a system from an initial macrostate, \textbf{I}, to a final macrostate, \textbf{II}, and finds an inequality which bounds from below the sum of the heat production and change in entropy by a quantity related to the transition probabilities between the two macrostates. This inequality, however, does not explicitly make use of a Bayesian update based on the observation of the system's final macrostate: as we have seen previously, the inclusion of Bayesian updates can significantly change one's interpretation of the entropy production. 

In seeking to interpret England's inequality within our framework, we consider the form of the BSL in an experiment where the initial distribution has support only on the initial macrostate, and the measurement at the conclusion determines the final macrostate. This is a slight generalization of the Boltzmann setup considered in Section~\ref{sec-specialcases} above. We then have the option to consider the difference between the entropy of the updated final distribution and the entropy of either the updated or unupdated initial distribution.

First, making use of the unupdated initial state, it can be shown that
\be \label{eq-England-Unupdated}
S(\rho_{\tau|\mathrm{\textbf{II}}}) -S(\rho_{0}) + \ev{\mathcal{Q}}_{F|\mathrm{\textbf{II}}} \geq - \log \frac{ \pi( \mathrm{\textbf{II}} \rightarrow \mathrm{\textbf{I}} ) }{ \pi( \mathrm{\textbf{I}} \rightarrow \mathrm{\textbf{II}} ) } + S(\rho_{0|\mathrm{\textbf{II}}}) - S(\rho_0).
\ee
This inequality is similar in spirit to England's: when $S(\rho_{0|\mathrm{\textbf{II}}}) \geq S(\rho_0)$, England's inequality immediately follows. Alternatively, using the updated initial state, we find
\be \label{eq-England-Updated}
S(\rho_{\tau|\mathrm{\textbf{II}}}) -S(\rho_{0|\mathrm{\textbf{II}}}) +  \ev{\mathcal{Q}}_{F|\mathrm{\textbf{II}}} \geq D(\rho_{0|\mathrm{\textbf{II}}}\| \tilde{\rho}_\mathrm{\textbf{II}}) + D(\rho_{0|\mathrm{\textbf{II}}}\|\rho_0)-D(\rho_{\tau|\mathrm{\textbf{II}}}\|\rho_\tau) \geq - \log \frac{\pi(\mathrm{\textbf{II}} \rightarrow \mathrm{\textbf{I}})}{\pi( \mathrm{\textbf{I}} \rightarrow \mathrm{\textbf{II}})}.
\ee
This differs from England's result only in that the entropy of the initial state has been replaced by the entropy of the updated initial state. Making this adjustment to England's inequality, we recover his bound from the bound given by the BSL. (We thank Timothy Maxwell for proving this relation.)

\paragraph{Future directions.} In this paper we have concentrated on incorporating Bayesian updates into the basic formalism of statistical mechanics, but a number of generalizations and applications present themselves as directions for future research. Potential examples include optimization of work-extraction (so-called ``Maxwell's demon" experiments) and cooling in nanoscale systems, as well as possible applications to biological systems.
It would be interesting to experimentally test the refined versions of the ordinary and Bayesian Second Laws, to quantify how close the inequalities are to being saturated. We are currently working to extend the BSL to quantum systems.

\section*{Acknowledgments} It is a pleasure to thank Gavin Crooks, Christopher Jarzynski, Timothy Maxwell, and Nicole Yunger Halpern for helpful conversations. 
This research is funded in part by the Walter Burke Institute for Theoretical Physics at Caltech, by DOE grant DE-SC0011632, and by the Gordon and Betty Moore Foundation through Grant 776 to the Caltech Moore Center for Theoretical Cosmology and Physics. The work of SL is supported in part by the Berkeley Center for Theoretical Physics, by the National Science Foundation (award numbers 1214644 and 1316783), by fqxi grant RFP3-1323, and by the US Department of Energy under Contract DE-AC02-05CH11231.

\newpage

\bibliographystyle{utphys}
\bibliography{BSLDraft-S1}

\newpage

\begin{appendix}
\section{Oscillator Evolution}
\label{app:Oscillator}
Here we show plots of the distribution functions for the three numerical harmonic-oscillator experiments discussed in Section \ref{sec:Oscillator}. 

\begin{figure}[!ht]
	\centering
	\begin{subfigure}{0.3\textwidth}
		\includegraphics[width=\textwidth]{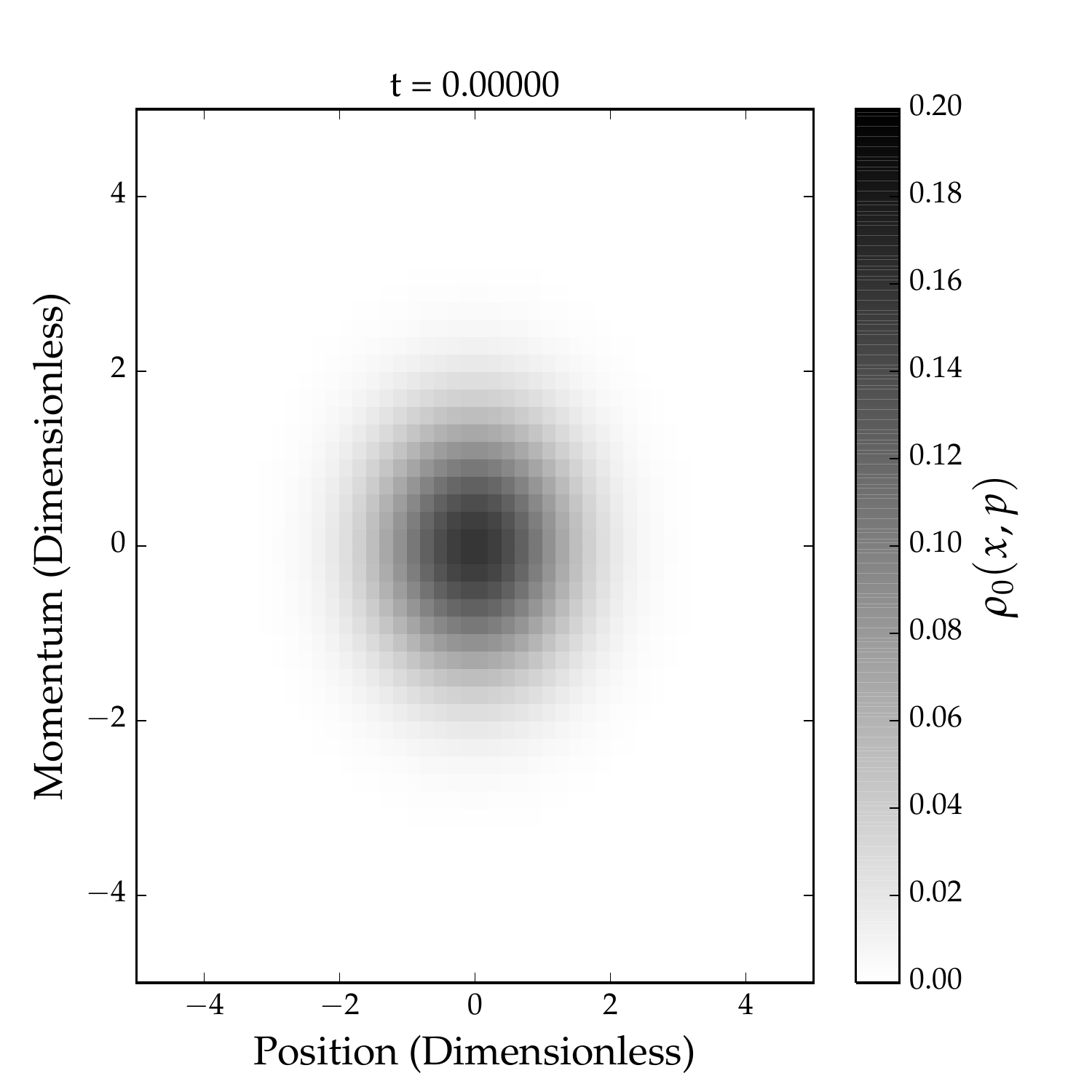}
		\caption{Initial distribution}
		\label{fig:initDist-static1}
	\end{subfigure}
	\begin{subfigure}{0.3\textwidth}
		\includegraphics[width=\textwidth]{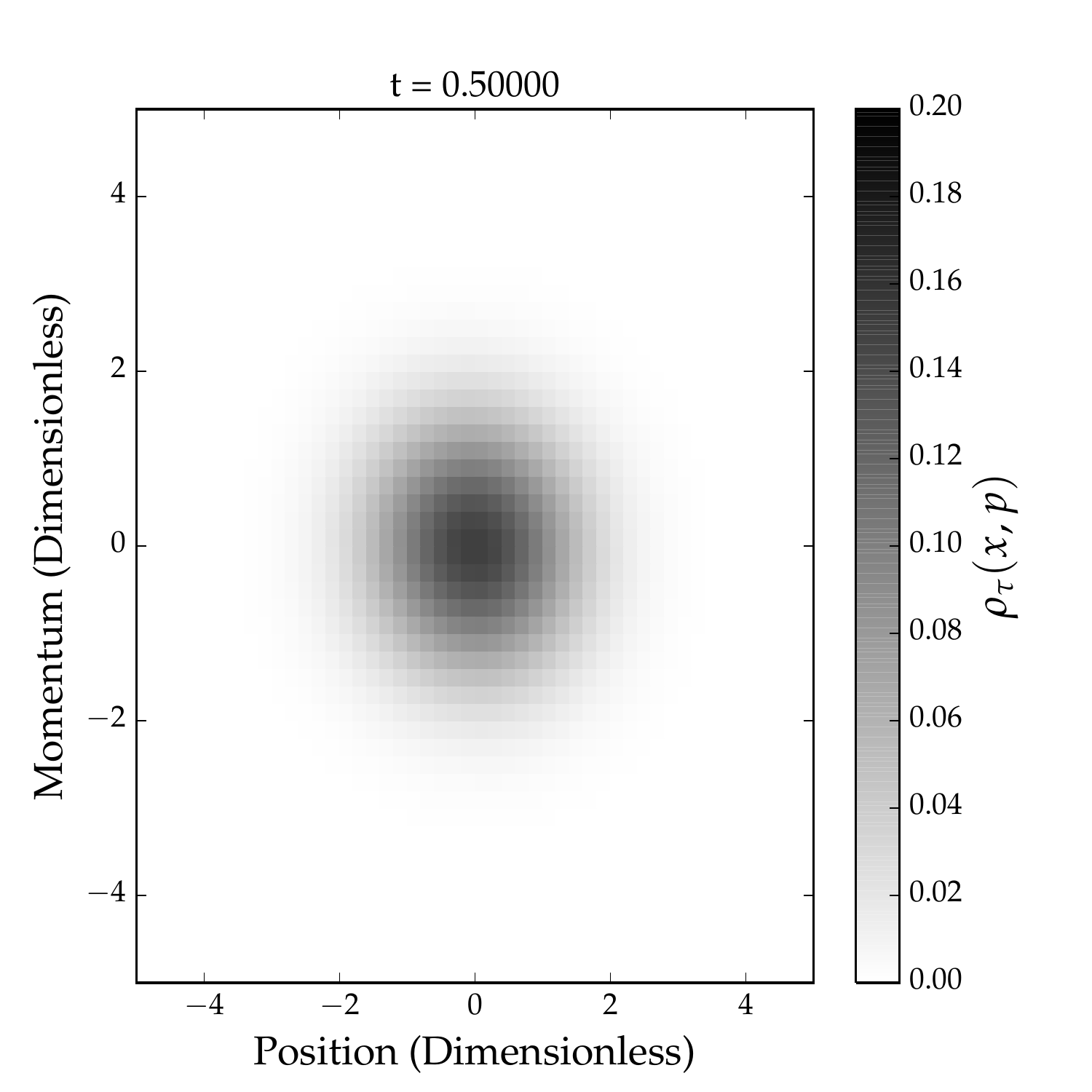}
		\caption{Final distribution}
		\label{fig:finDist-static1}
	\end{subfigure}
	\begin{subfigure}{0.3\textwidth}
		\includegraphics[width=\textwidth]{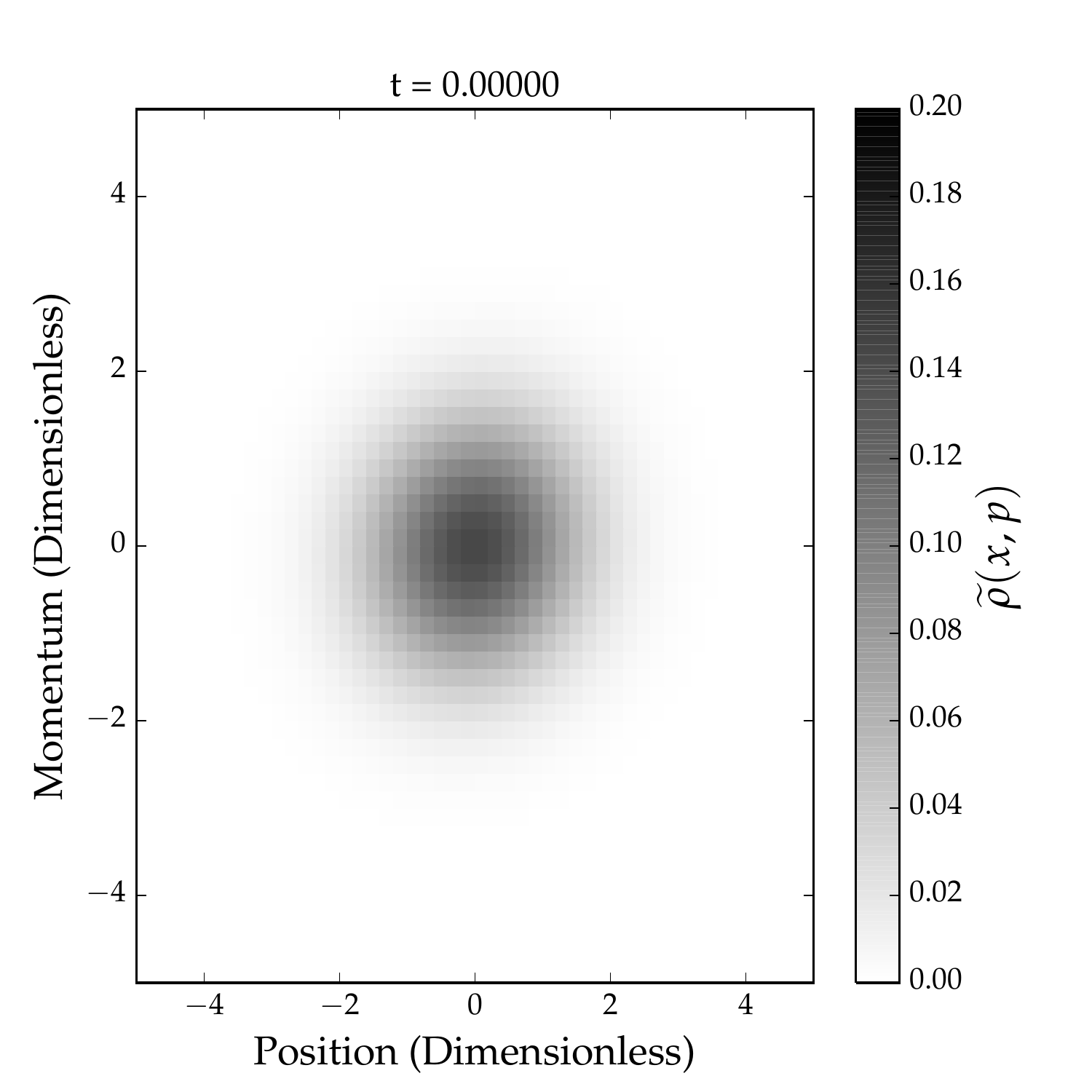}
		\caption{Cycled distribution}
		\label{fig:cycledDist-static1}
	\end{subfigure}
	\begin{subfigure}{0.3\textwidth}
		\includegraphics[width=\textwidth]{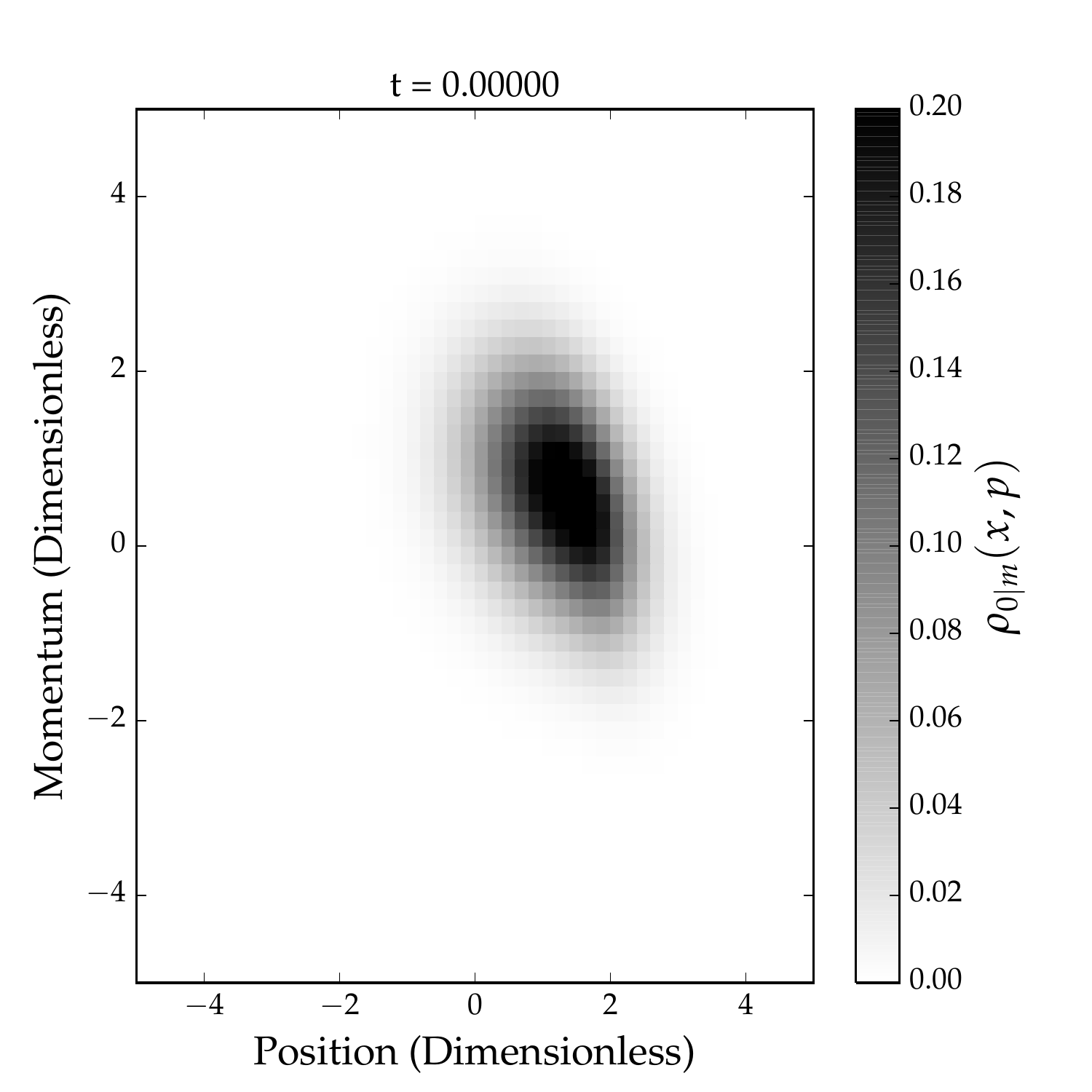}
		\caption{Updated initial \newline distribution}
		\label{fig:updatedInitDist-static1}
	\end{subfigure}
	\begin{subfigure}{0.3\textwidth}
		\includegraphics[width=\textwidth]{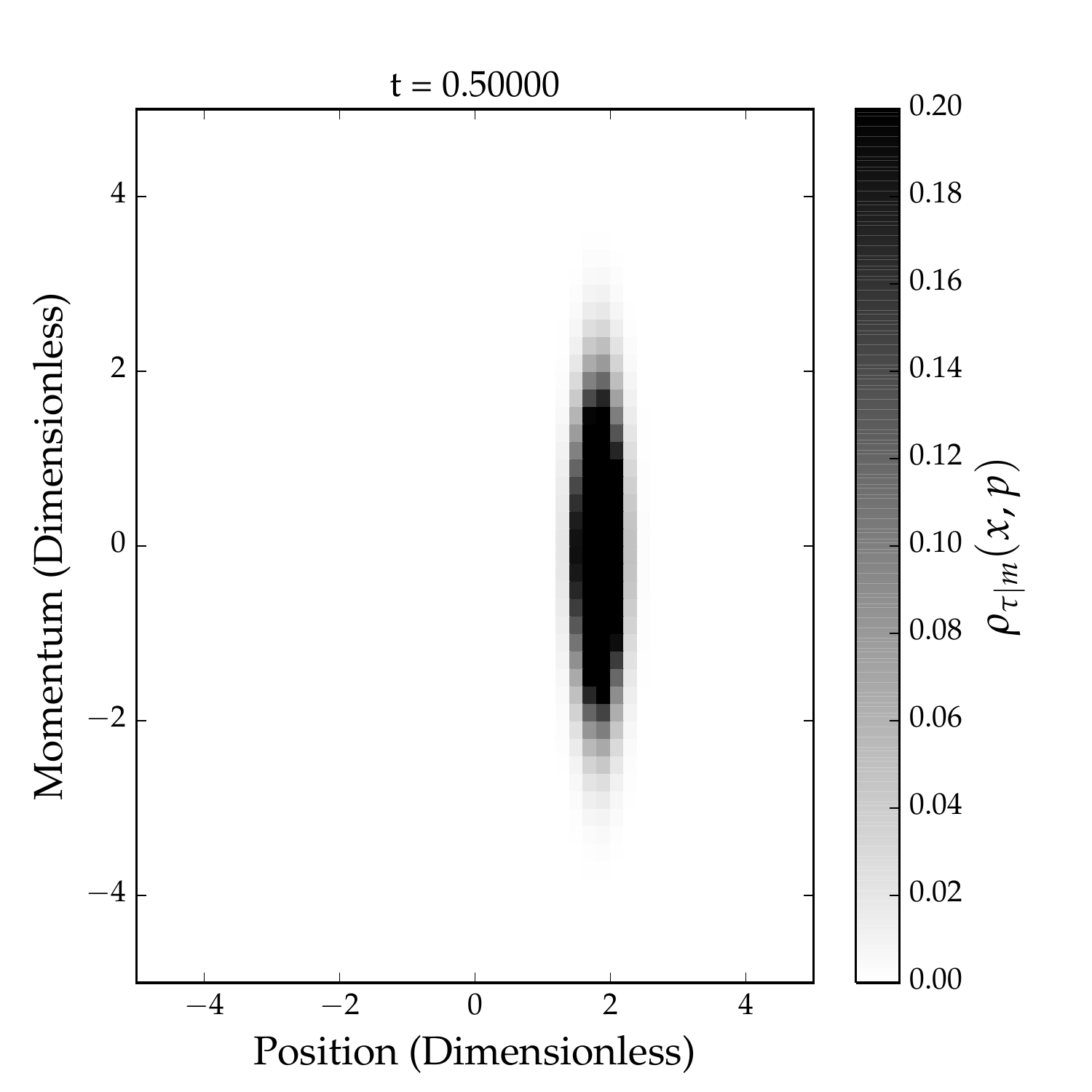}
		\caption{Updated final \newline distribution}
		\label{fig:updatedFinDist-static1}
	\end{subfigure}
	\begin{subfigure}{0.3\textwidth}
		\includegraphics[width=\textwidth]{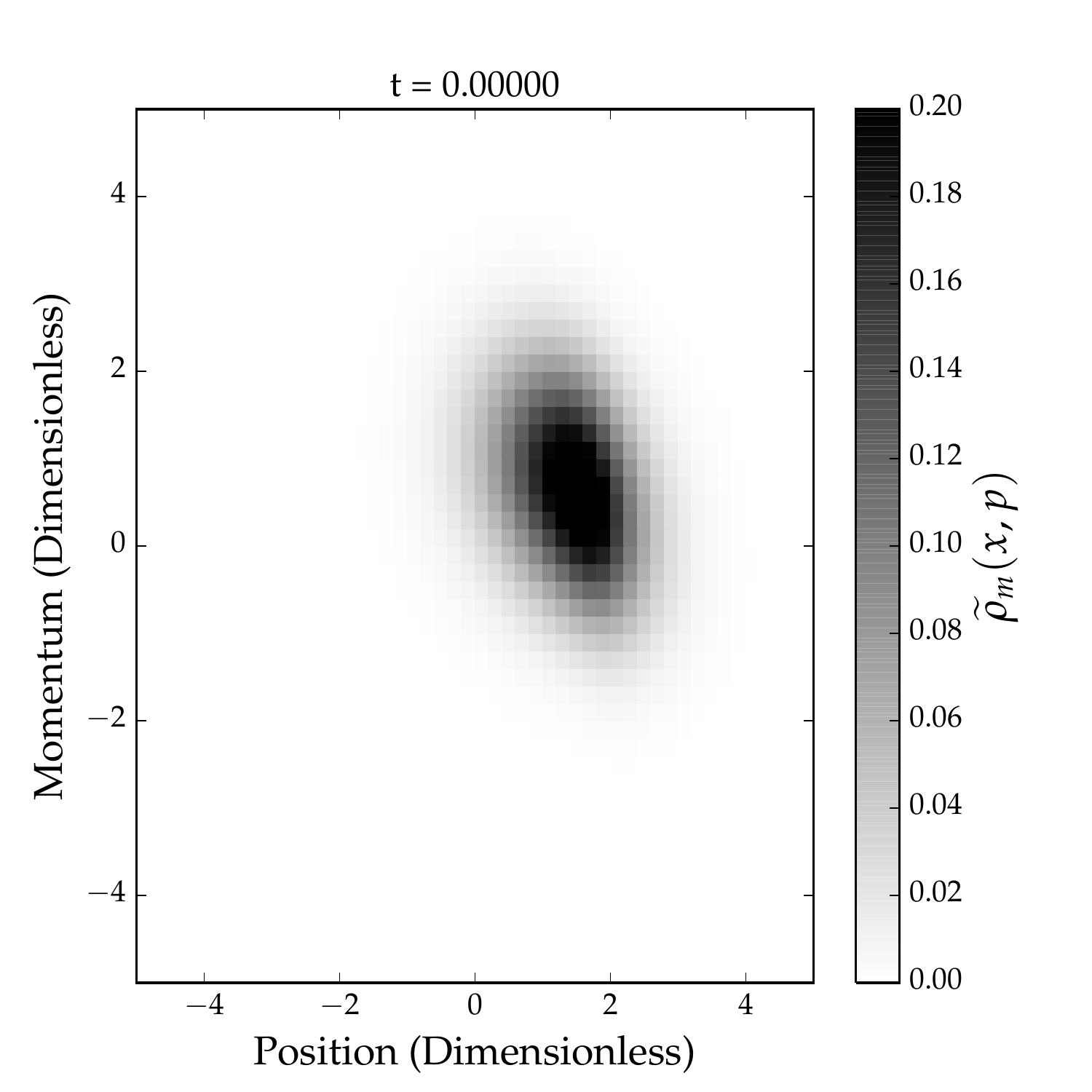}
		\caption{Updated cycled \newline distribution}
		\label{fig:updatedCycledDist-static1}
		\label{fig:other}
	\end{subfigure}
	\caption{Evolution of a damped harmonic oscillator coupled to a heat bath in initial thermal equilibrium under a trivial protocol. Units are chosen such that $M=1$, $k(t=0)=1$, and $\beta=1$. Each graph shows the phase space probability distribution with respect to position and momentum at different points in the experiment.}
	\label{fig:evo-static1}
\end{figure}

\begin{figure}[!ht]
	\centering
	\begin{subfigure}{0.3\textwidth}
		\includegraphics[width=\textwidth]{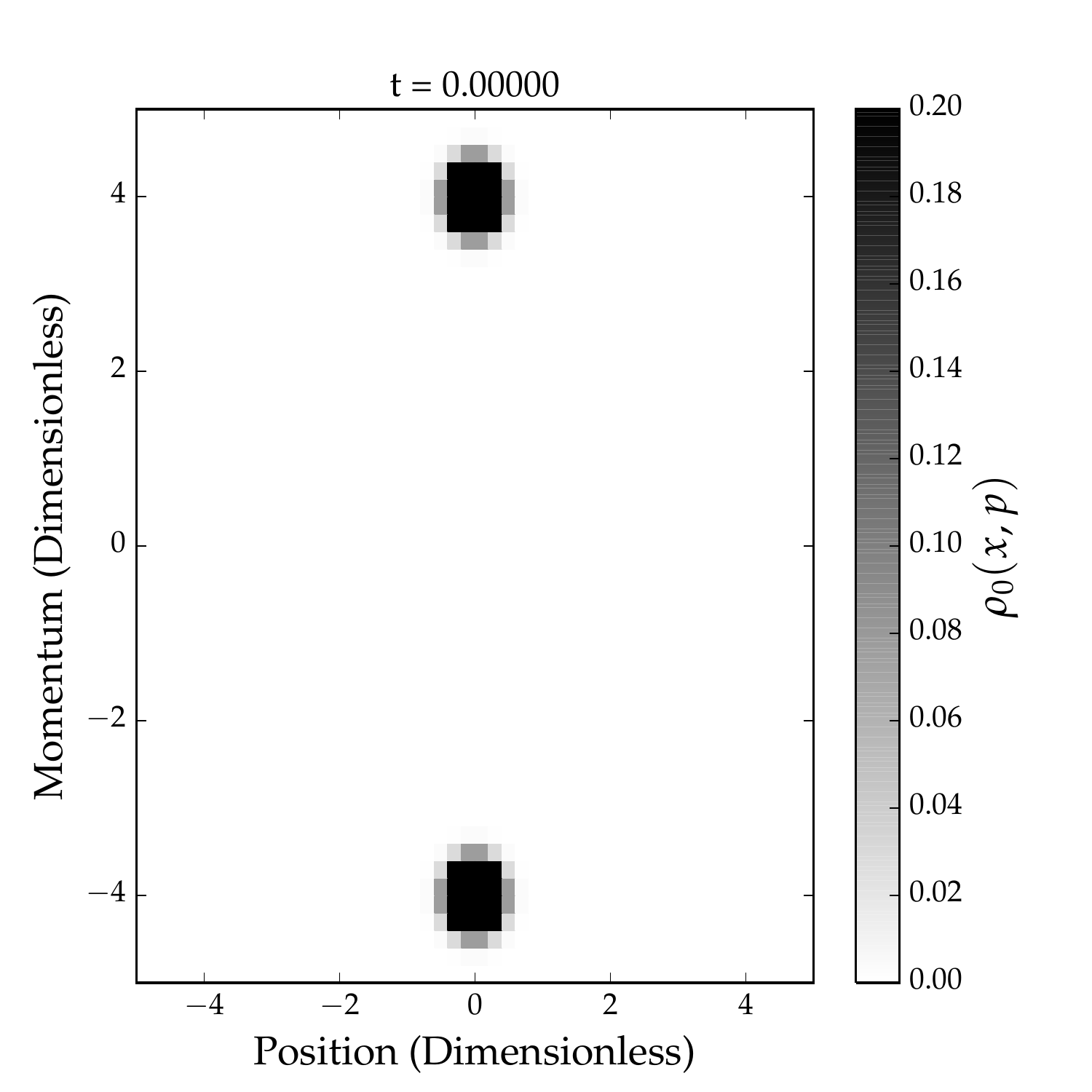}
		\caption{Initial distribution}
		\label{fig:initDist-static2}
	\end{subfigure}
	\begin{subfigure}{0.3\textwidth}
		\includegraphics[width=\textwidth]{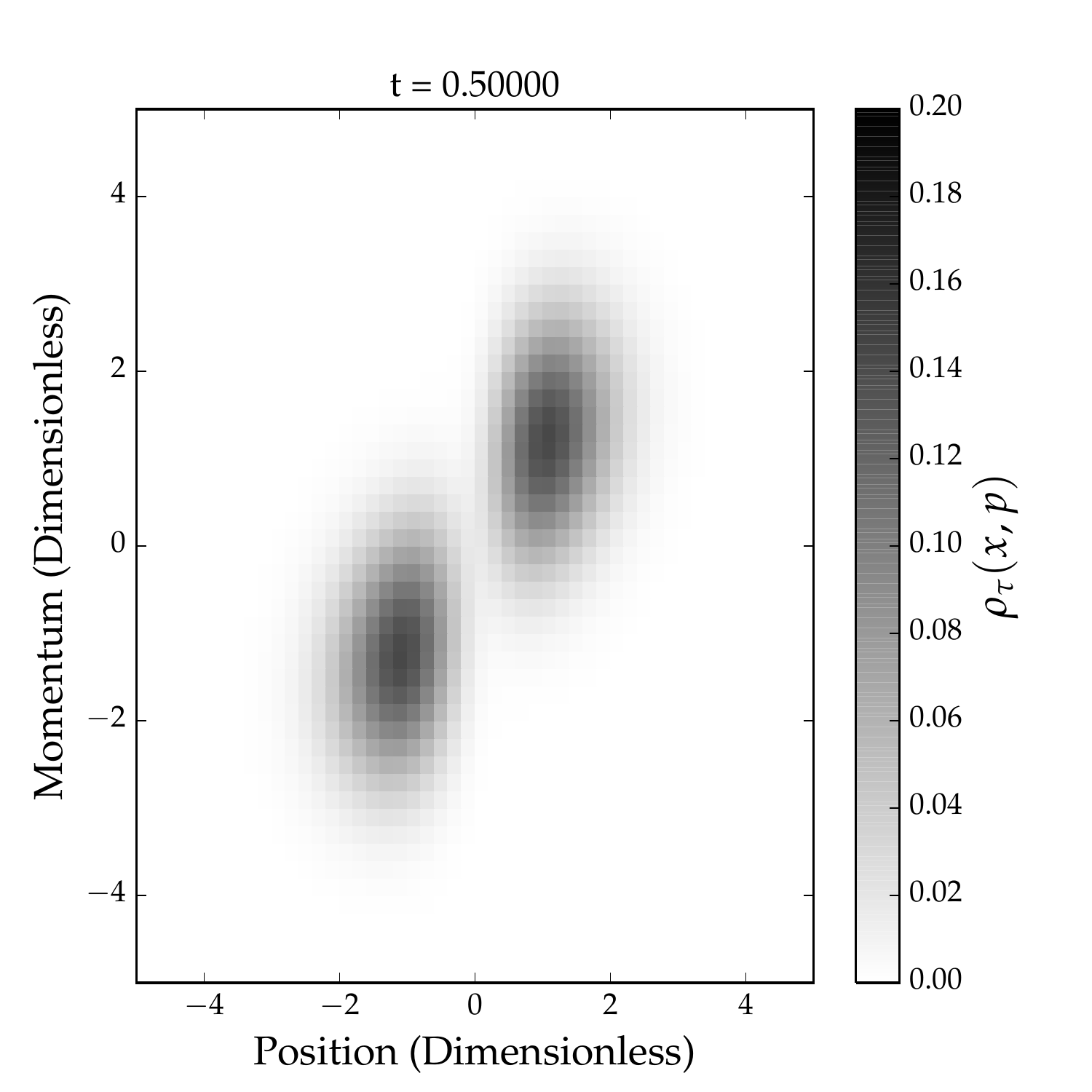}
		\caption{Final distribution}
		\label{fig:finDist-static2}
	\end{subfigure}
	\begin{subfigure}{0.3\textwidth}
		\includegraphics[width=\textwidth]{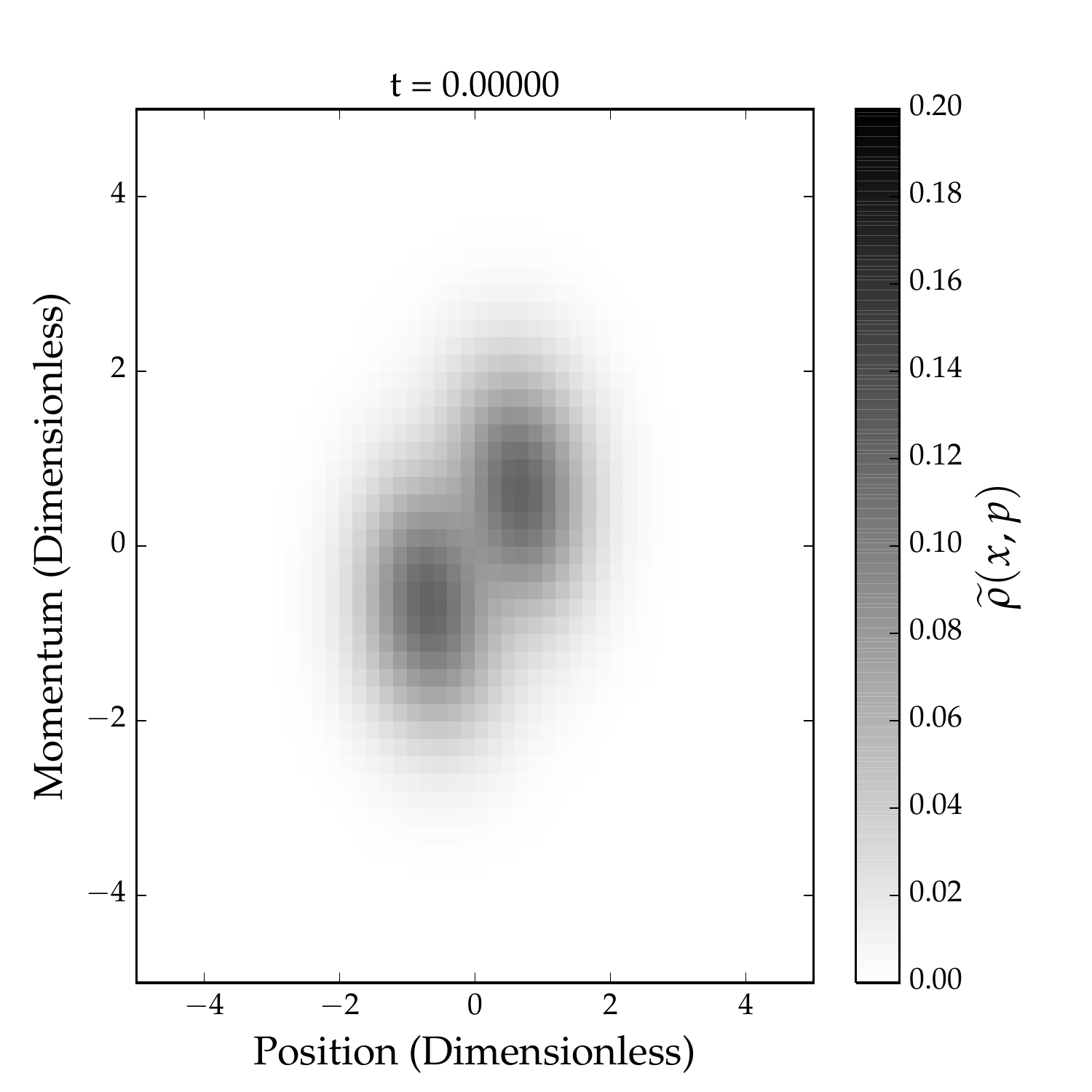}
		\caption{Cycled distribution}
		\label{fig:revProFinDist-static2}
	\end{subfigure}
	\begin{subfigure}{0.3\textwidth}
		\includegraphics[width=\textwidth]{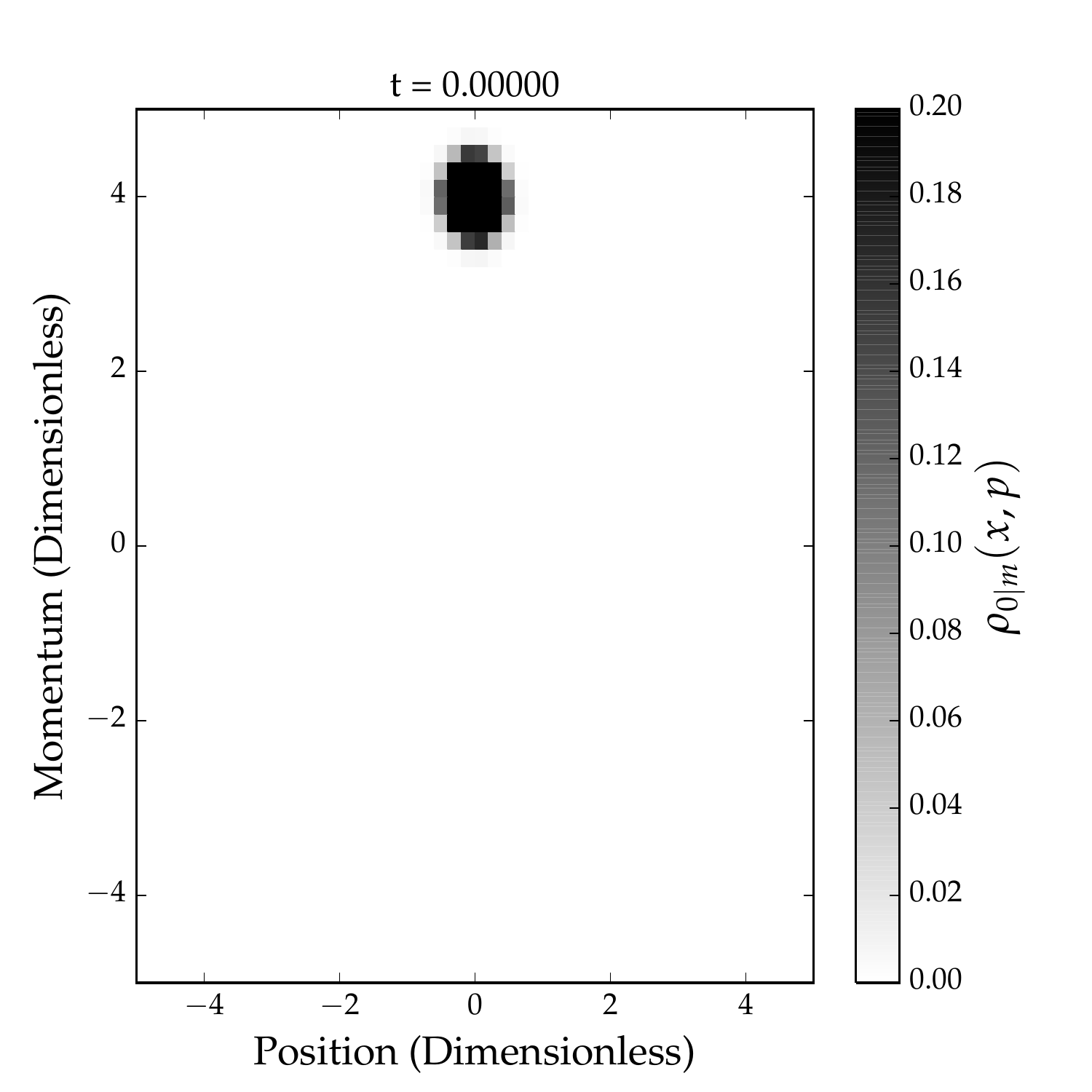}
		\caption{Updated initial \newline distribution}
		\label{fig:updatedInitDist-static2}
	\end{subfigure}
	\begin{subfigure}{0.3\textwidth}
		\includegraphics[width=\textwidth]{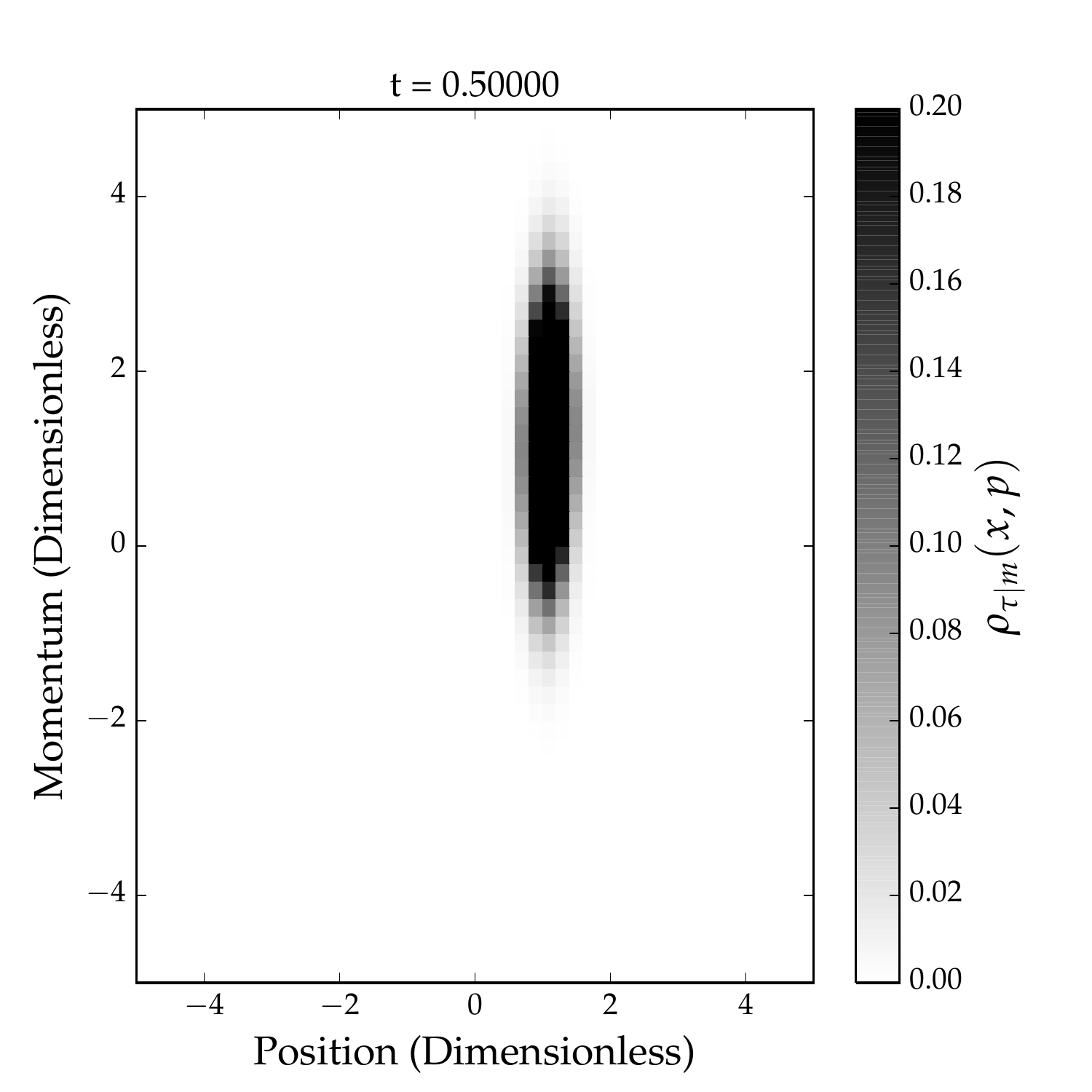}
		\caption{Updated final \newline distribution}
		\label{fig:updatedFinDist-static2}
	\end{subfigure}
	\begin{subfigure}{0.3\textwidth}
		\includegraphics[width=\textwidth]{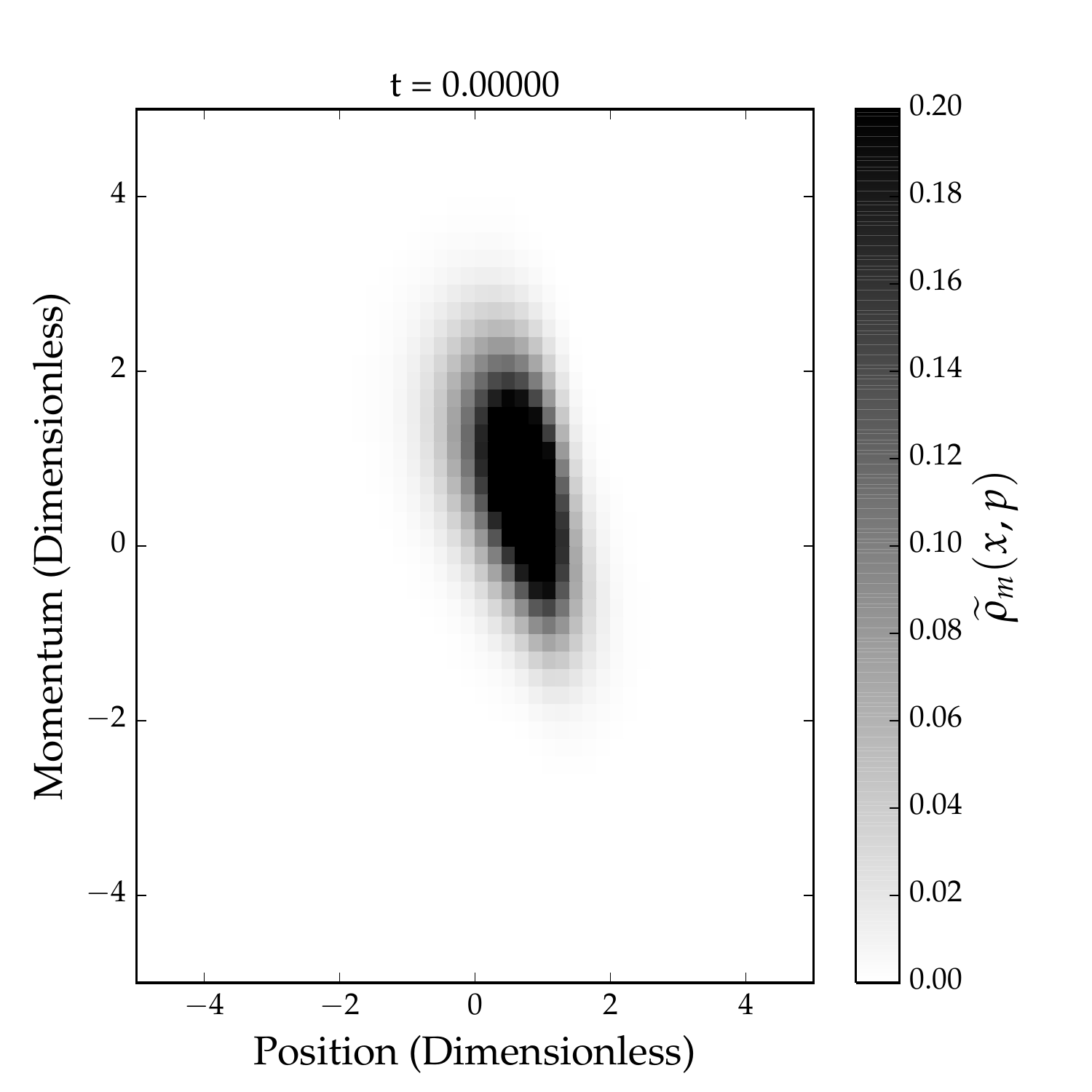}
		\caption{Updated cycled \newline distribution}
		\label{fig:conjFinDist-static2}
	\end{subfigure}
	\caption{Evolution of a damped harmonic oscillator coupled to a heat bath with known position and magnitude of momentum under a trivial protocol. Units are chosen such that $M=1$, $k(t=0)=1$, and $\beta=1$. Each graph shows the phase space probability distribution with respect to position and momentum at different points in the experiment.}
	\label{fig:evo-static2}
\end{figure}

\begin{figure}[!ht]
	\centering
	\begin{subfigure}{0.3\textwidth}
		\includegraphics[width=\textwidth]{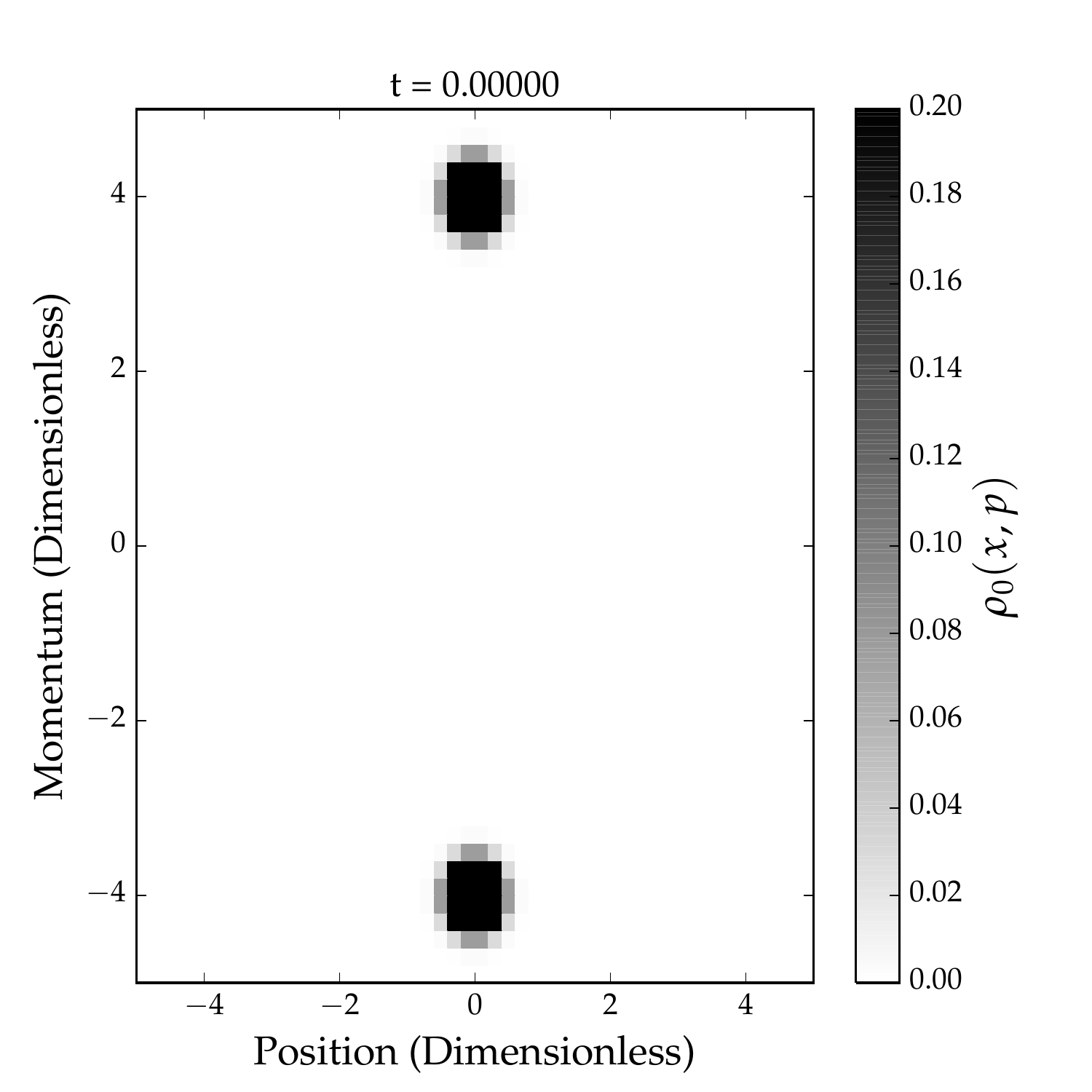}
		\caption{Initial distribution}
		\label{fig:initDist-drag}
	\end{subfigure}
	\begin{subfigure}{0.3\textwidth}
		\includegraphics[width=\textwidth]{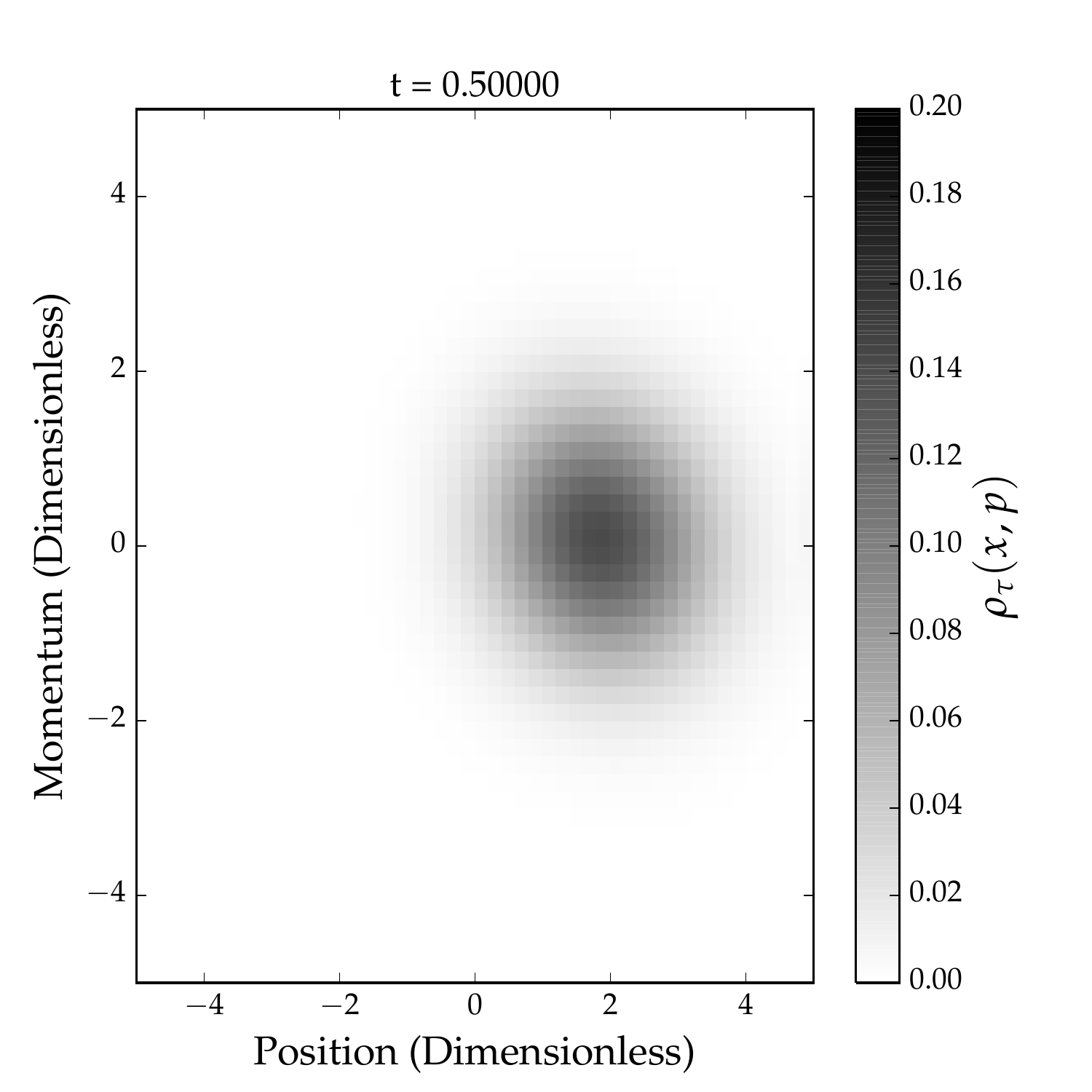}
		\caption{Final distribution}
		\label{fig:finDist-drag}
	\end{subfigure}
	\begin{subfigure}{0.3\textwidth}
		\includegraphics[width=\textwidth]{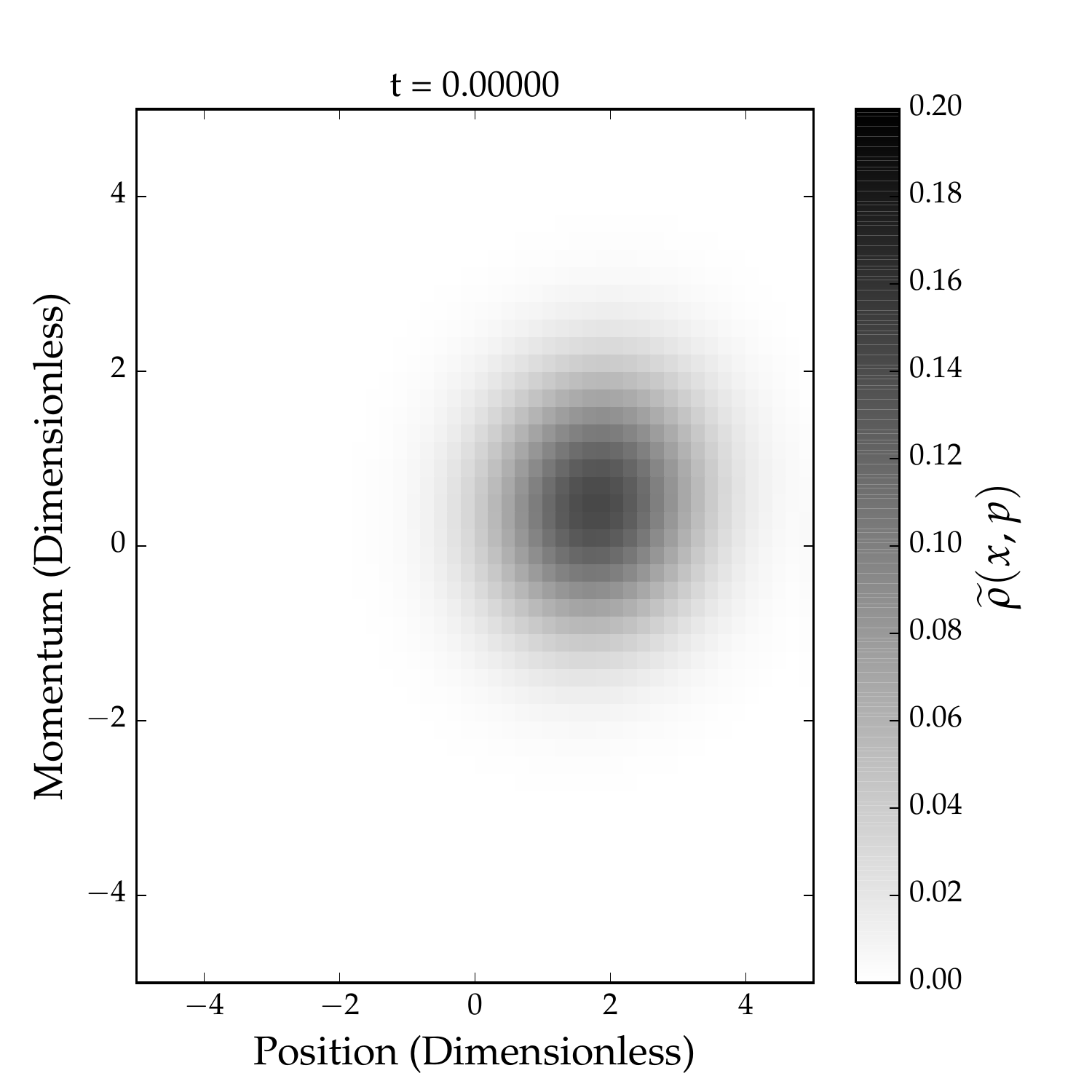}
		\caption{Cycled distribution}
		\label{fig:revProFinDist-drag}
	\end{subfigure}
	\begin{subfigure}{0.3\textwidth}
		\includegraphics[width=\textwidth]{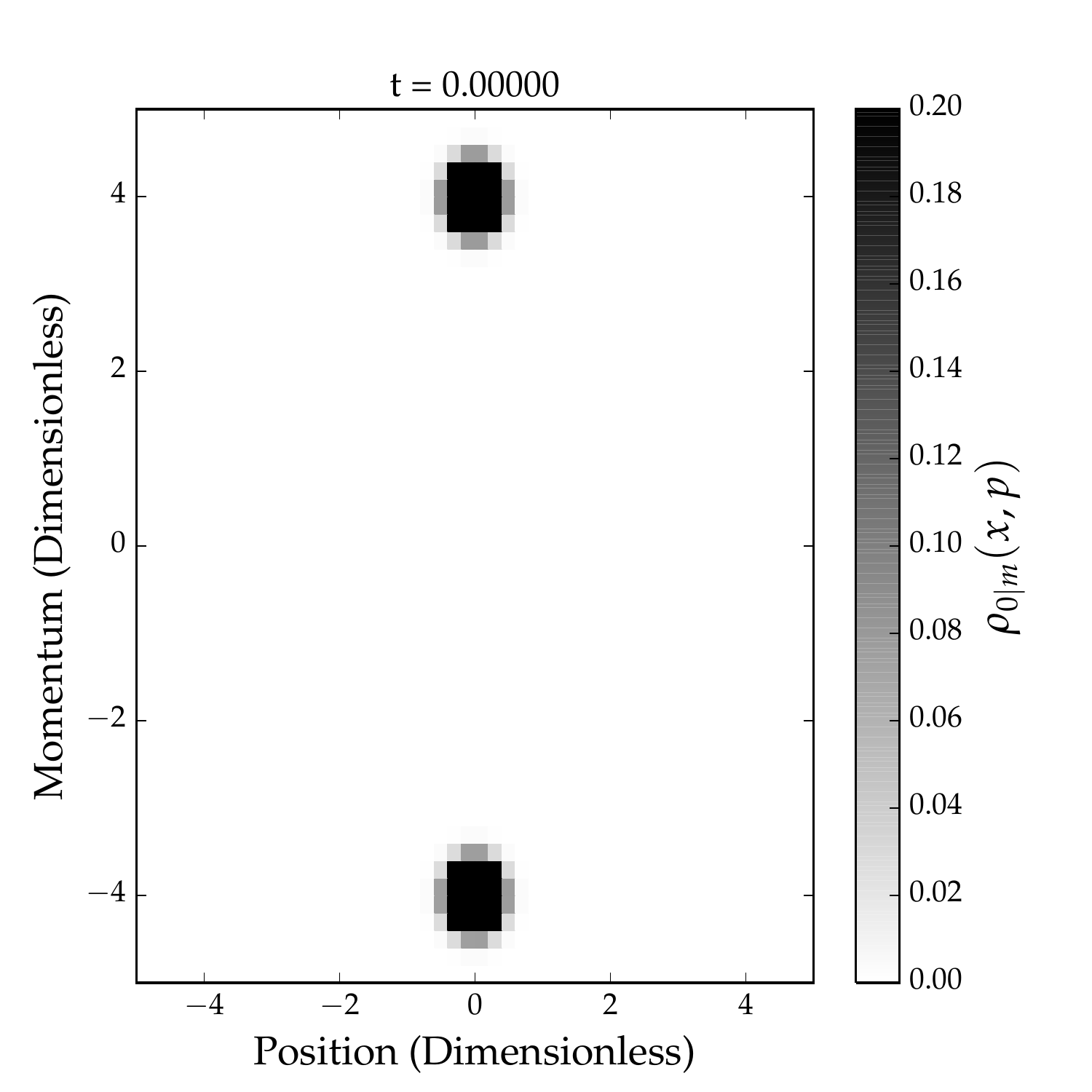}
		\caption{Updated initial \newline distribution}
		\label{fig:updatedInitDist-drag}
	\end{subfigure}
	\begin{subfigure}{0.3\textwidth}
		\includegraphics[width=\textwidth]{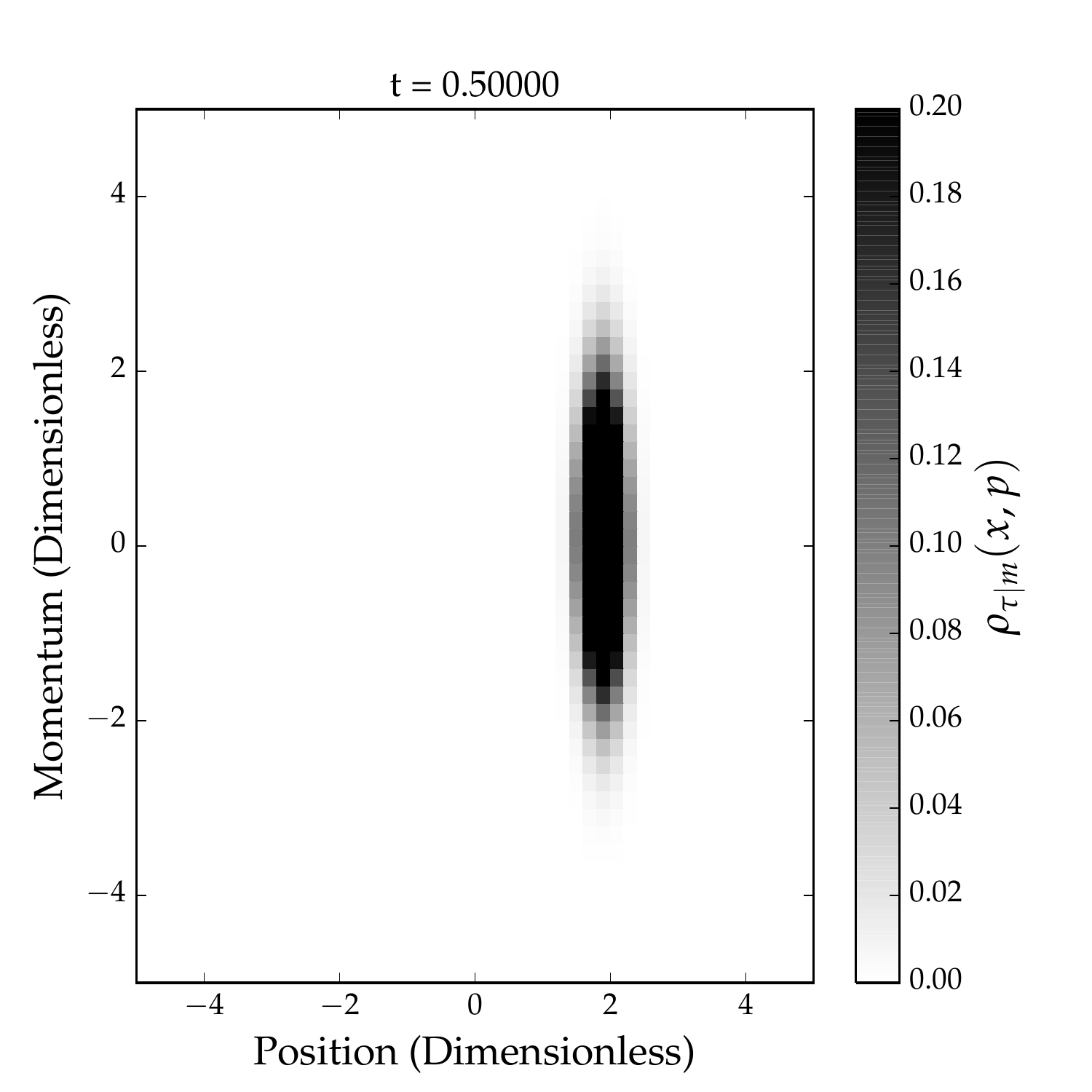}
		\caption{Updated final \newline distribution}
		\label{fig:updatedFinDist-drag}
	\end{subfigure}
	\begin{subfigure}{0.3\textwidth}
		\includegraphics[width=\textwidth]{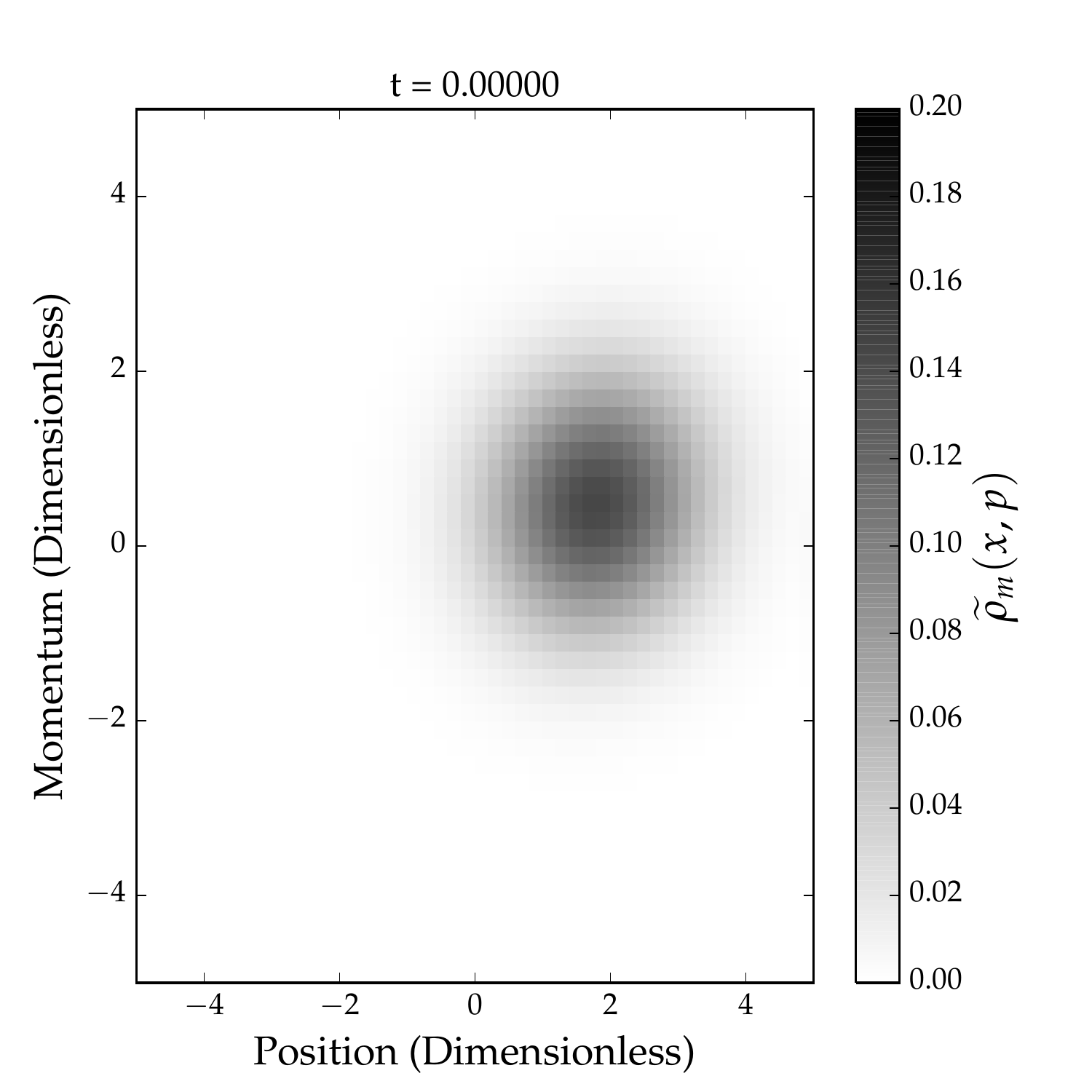}
		\caption{Updated cycled \newline distribution}
		\label{fig:conjFinDist-drag}
	\end{subfigure}
	\caption{Evolution of a damped harmonic oscillator coupled to a heat bath in initial thermal equilibrium under a ``dragging'' protocol. Units are chosen such that $M=1$, $k(t=0)=1$, and $\beta=1$. Each graph shows the phase space probability distribution with respect to position and momentum at different points in the experiment.}
	\label{fig:evo-drag}
\end{figure}

\end{appendix}

\end{document}